\documentclass[11pt,document,nofootinbib,superscriptaddress,onecolumn,preprintnumbers,balancelastpage]{article}

\usepackage{jheppub}

\usepackage{subfig}
\usepackage[countmax]{subfloat}
\usepackage{framed}
\usepackage{mdframed}
\usepackage{multirow}
\usepackage{slashed}
\usepackage{booktabs} 
\usepackage{xcolor}

\usepackage{array}

\makeatletter  
\newcommand{\thickhline}{%
    \noalign {\ifnum 0=`}\fi \hrule height 1pt
    \futurelet \reserved@a \@xhline
}

\newcolumntype{"}{@{\hskip\tabcolsep\vrule width 1pt\hskip\tabcolsep}}

\makeatother
\def\beq{\begin{eqnarray}}
\def\eeq{\end{eqnarray}}
\def\bea{\begin{eqnarray}}
\def\eea{\end{eqnarray}}
\usepackage{latexsym}
\usepackage{amssymb}
\usepackage{amsmath}
\usepackage{epsfig,amsmath,graphics}
\usepackage{listings}
\usepackage{color}
\usepackage{dcolumn}
\usepackage{slashed}
\usepackage{cancel}
\usepackage{comment,latexsym} 
\usepackage{bm}
\usepackage{verbatim}
\usepackage{tabularx}
\usepackage{dcolumn}

\usepackage{setspace}
\usepackage{xspace}
\usepackage{multirow}
\usepackage{titlesec}
\usepackage[bbgreekl]{mathbbol}
\usepackage{bm}
\usepackage{float}
\usepackage{placeins}
\usepackage{afterpage}
\usepackage{footmisc}
\usepackage{breakcites}

\usepackage{empheq}

\usepackage[toc,page]{appendix}

\usepackage{etoolbox}
\appto\appendix{\addtocontents{toc}{\protect\setcounter{tocdepth}{1}}}

\usepackage{scalerel}

\usepackage{arydshln}

\allowdisplaybreaks

\expandafter\def\expandafter\normalsize\expandafter{%
    \normalsize
    \setlength\abovedisplayskip{8pt}
    \setlength\belowdisplayskip{8pt}
    \setlength\abovedisplayshortskip{8pt}
    \setlength\belowdisplayshortskip{8pt}
}

\definecolor{darkblue}{rgb}{0,0,0.5}

\definecolor{colorTC}{rgb}{.2,.7,.2}

\newcommand{\be}{\begin{equation}}
\newcommand{\ee}{\end{equation}}

\newcommand{\gsim}{\lower.7ex\hbox{$\;\stackrel{\textstyle>}{\sim}\;$}}
\newcommand{\lsim}{\lower.7ex\hbox{$\;\stackrel{\textstyle<}{\sim}\;$}}

\newcommand{\nocontentsline}[3]{}
\newcommand{\tocless}[2]{\bgroup\let\addcontentsline=\nocontentsline#1{#2}\egroup}

\title{
Baryogenesis and first-order QCD transition with gravitational waves from a large lepton asymmetry}
\author[a]{Fei Gao,}
\author[b]{Julia Harz,}
\author[c]{Chandan Hati,}
\author[d]{Yi Lu,}
\author[e,f]{Isabel M. Oldengott,}
\author[g]{Graham White}
\emailAdd{fei.gao@bit.edu.cn}
\emailAdd{julia.harz@uni-mainz.de}
\emailAdd{chandan@ific.uv.es}
\emailAdd{qwertylou@pku.edu.cn}
\emailAdd{ioldengott@physik.uni-bielefeld.de}
\emailAdd{g.a.white@soton.ac.uk}
\affiliation[a]{ School of Physics, Beijing Institute of Technology, 100081 Beijing, China}
\affiliation[b]{PRISMA$^+$ Cluster of Excellence \& Mainz Institute for Theoretical Physics, FB 08 - Physics, Mathematics and Computer
Science, Johannes Gutenberg-Universit\"{a}t Mainz, 55099 Mainz, Germany}
\affiliation[c]{ Instituto de F\'{i}sica Corpuscular (IFIC), Universitat de Val\`encia-CSIC, C/ Catedratico Jose Beltran, 2, E-46980 Valencia, Spain}
\affiliation[d]{ Department of Physics and State Key Laboratory of Nuclear Physics and Technology, Peking University, Beijing 100871, China}
\affiliation[e]{  Centre for Cosmology, Particle Physics and Phenomenology, Universit\'e catholique de Louvain, Chemin du cyclotron, 2, Louvain-la-Neuve B-1348, Belgium}

\affiliation[f]{  Fakultät für Physik, Bielefeld University, D–33501 Bielefeld, Germany}
\affiliation[g]{ School of Physics and Astronomy, University of Southampton, Southampton SO17 1BJ, United Kingdom}

\abstract{A large primordial lepton asymmetry can lead to successful baryogenesis by preventing the restoration of electroweak symmetry at high temperatures, thereby suppressing the sphaleron rate. This asymmetry can also lead to a first-order cosmic QCD transition, accompanied by detectable gravitational wave (GW) signals. By employing next-to-leading order dimensional reduction we determine that the necessary lepton asymmetry is approximately one order of magnitude smaller than previously estimated. Incorporating an updated QCD equation of state that harmonizes lattice and functional QCD outcomes, we pinpoint the range of lepton flavor asymmetries capable of inducing a first-order cosmic QCD transition. To maintain consistency with observational constraints from the Cosmic Microwave Background and Big Bang Nucleosynthesis, achieving the correct baryon asymmetry requires entropy dilution by approximately a factor of ten. However, the first-order QCD transition itself can occur independently of entropy dilution. We propose that the sphaleron freeze-in mechanism can be investigated through forthcoming GW experiments such as $\mu$Ares.}

\begin{document} 
\preprint{\hfill MITP-24-060}
\maketitle
\flushbottom

\section{Introduction}
One of the biggest open puzzles of particle physics and cosmology is the origin of the matter-antimatter asymmetry of the Universe. Measurements of primordial element abundances and the Cosmic Microwave Background (CMB) reveal that the Universe has a tiny asymmetry in baryons, at the order of $\mathcal{O}(10^{-11})$. According to the Standard Model (SM) of particle physics, this baryon asymmetry should however be smaller by many orders of magnitudes. Explaining the creation of the observed baryon asymmetry necessitates theories of baryogenesis and leptogenesis. 

In standard cosmology (based on the most common baryogenesis via leptogenesis mechanisms) the lepton flavor asymmetries are assumed to be at the same order of magnitude as the baryon asymmetry. Large lepton asymmetries can however significantly alter the cosmic evolution at various epochs: they could induce hypermagnetic fields~\cite{Domcke:2022uue}, change the decoupling of WIMPS~\cite{Stuke:2011wz}, modify the sphaleron conversion rate~\cite{McDonald:1999in,March-Russell:1999hpw,Barenboim:2017dfq}, have an impact on the epoch of quantum chromodynamics (QCD)~\cite{Zarembo:2000wj,Schwarz:2009ii,Wygas:2018otj,Middeldorf-Wygas:2020glx,Gao:2021nwz}, change the production of primordial black holes~\cite{Bodeker:2020stj,Vovchenko:2020crk}, modify the inflationary gravitational waves (GWs) spectrum~\cite{Hajkarim:2019csy}, lead to the formation of a pion condensate in the early Universe~\cite{Vovchenko:2020crk,Middeldorf-Wygas:2020glx}, impact oscillations between active and sterile neutrinos~\cite{Shi:1998km}, change neutrino decoupling and the oscillations of active neutrinos~\cite{Dolgov:2002ab,Wong:2002fa,Pastor:2008ti,Mangano:2010ei,Mangano:2011ip,Barenboim:2016shh,Johns:2016enc,Froustey:2021azz}, modify big bang nucleosynthesis (BBN) \cite{Beaudet:1976,Steigman:2007xt} and the formation of the cosmic microwave background (CMB) \cite{Lesgourgues:1999wu}. 

Here, we study a scenario where large lepton flavor asymmetries themselves give rise to the observed value of the baryon asymmetry~\cite{McDonald:1999in,March-Russell:1999hpw,Barenboim:2017dfq}. Due to the non-restoration of the electroweak symmetry, the sphaleron rate gets suppressed in the presence of sufficiently large lepton asymmetries. Therefore the existence of a large primordial lepton asymmetry can provide a mechanism of baryogenesis, where the observed small baryon asymmetry of the Universe is generated via conversion of a large lepton asymmetry, if the sphaleron conversion rate is sufficiently slow due to the electroweak symmetry non-restoration. We refer to this baryogenesis scenario as \textit{sphaleron freeze-in}. While former works~\cite{Bajc:1997ky,Barenboim:2017dfq} applied the perturbative effective potential for baryons and leptons, we here use the technique of dimensional reduction~\cite{Kajantie:1995dw} to consistently take into account infrared divergences at finite temperature. Our calculations~\cite{Gao:2021nwz} show that the necessary lepton asymmetries required for successful baryogenesis are an order of magnitude smaller than it was estimated before. This improvement reduces the amount of entropy dilution required to produce the correct baryon asymmetry of the Universe, given the stringent constraints from Big Bang nucleosynthesis (BBN) and the cosmic microwave background (CMB) on the lepton asymmetry, which is predominantly stored in the form of Dirac neutrino flavour asymmetries at the time of CMB.

Interestingly, it was shown recently~\cite{Gao:2021nwz} that large lepton flavor asymmetries can also induce a first-order cosmic QCD transition. As a first-order phase transition is expected to be accompanied by the emission of gravitational waves (GWs), the new era of GW measurements would therefore allow to constrain the lepton asymmetries and potentially test the sphaleron freeze-in paradigm. The work of~\cite{Gao:2021nwz} is based on results from functional QCD, namely solutions of the Schwinger-Dyson equations within the so-called Rainbow-ladder truncation~\cite{Gao:2015kea}. While Ref.~\cite{Gao:2021nwz} depicts the first proof-of-principle of the possibility of a first-order cosmic QCD transition induced by large lepton flavor asymmetries it was also mentioned in~\cite{Gao:2021nwz} that the location of the QCD critical end point (CEP) from the Rainbow-ladder truncation was not predicted accurately. However improved truncation schemes and using results from the functional renormalization group has now made it possible to predict the location of the CEP more accurately and consistently~\cite{Gao:2020fbl,Gunkel:2021oya,Fu:2019hdw}. Very recently the thermodynamic quantities of QCD matter were calculated from an improved truncation scheme and by application of an Ising parameterization~\cite{Lu:2023msn}. In comparison to the method applied in~\cite{Gao:2021nwz}, this method is based on a more realistic position of the CEP and shows agreement with the results of lattice QCD at low chemical potentials. We here include the thermodynamic quantities calculated with the method of \cite{Lu:2023msn} into our calculation of the cosmic trajectory during the QCD epoch.

Our work addresses the following main questions: How large should the lepton asymmetries be in order to lead to successful baryogenesis via sphaleron freeze-in? Which values of the lepton asymmetries would induce a first-order cosmic QCD transition? How large is the expected GW signal from a first-order QCD transition? Finally, the main and all-embracing question is whether the sphaleron freeze-in mechanism is testable with future GW detectors. 

The structure of this work is as follows: In sec.~\ref{sec:The lepton asymmetry of our Universe} we discuss current constraints on the lepton asymmetries. Sec.~\ref{sec:Symmetry non-restoration and sphaleron freeze-in} is devoted to the sphaleron freeze-in mechanism and a discussion of an epoch of late-time entropy dilution which is needed for successful baryogenesis (consistent with observational constraints on the lepton asymmetries discussed in sec.~\ref{sec:The lepton asymmetry of our Universe}). In sec.~\ref{sec:Cosmic QCD epoch} we discuss the new functional QCD technique applied in this work and its impact on our calculations concerning the cosmic QCD epoch in the presence of large lepton asymmetries. Sec.~\ref{sec:Gravitational waves} addresses the expected GWs from a first-order QCD transition and its detection prospects. We conclude in sec.~\ref{sec:Conclusions}.


\section{The lepton asymmetry of our Universe}
\label{sec:The lepton asymmetry of our Universe}
The baryon asymmetry of the Universe can be defined as
\begin{equation}
    Y_B    =  \frac{n_B}{s}=\sum_{i}\frac{b_i n_i}s \, ,\label{eq:baryon_asymmetry}
\end{equation}
where the sum goes over all particle species carrying a baryon number, and $s$ is the total entropy density of the Universe. Here, $b_i$ denotes the baryon number and $n_i$ the net number density (particle minus anti-particle) of particle species $i$. Note that we assume the neutrinos to be Dirac such that this definition can be applied.  From measurements of 
primordial element abundances and the CMB, we know that   $Y_B = (8.70 \pm 0.06) \times 10^{-11}$ (inferred from \cite{Aghanim:2018eyx}).

Analogously to the baryon asymmetry in Eq. \eqref{eq:baryon_asymmetry}, the lepton asymmetry can be defined as the sum over individual lepton flavor asymmetries $Y_{L_{\alpha}}$,
\begin{equation}
    Y_{L} = \sum_{\alpha} Y_{L_{\alpha}} = \sum_{\alpha} \frac{n_{\alpha} + n_{\nu_{\alpha}}}{s} \, ,
    \label{eq:lepton_asymmetry}
\end{equation}
where $\alpha = e, \mu, \tau$. Note here the number density for neutrinos is taken also with the Fermi-Dirac distribution. 
The idea behind leptogenesis is to create an asymmetry in the leptonic sector which is converted to a baryon asymmetry via sphaleron processes. This then leads to the standard relation $Y_B = -\frac{28}{51} Y_L$ \cite{Harvey:1990qw}, implying a lepton asymmetry at the same order of magnitude as the baryon asymmetry. However, this standard relation only holds under the assumption that sphaleron processes are sufficiently efficient. As we explicitly demonstrate in sec.~\ref{sec:Symmetry non-restoration and sphaleron freeze-in}, for large lepton asymmetries sphaleron processes get suppressed such that the observed baryon asymmetry could counter-intuitively also originate from large lepton asymmetries.

In the following, we discuss the constraints on the total lepton asymmetry $Y_L$ as well as on the individual lepton flavor asymmetries $Y_{L_{\alpha}}$.   

\subsection{BBN and CMB bounds on lepton asymmetry}\label{sec:BBNandCMBconstraints}
As a consequence of the electric charge neutrality of the Universe, today the asymmetry in charged leptons must compensate for the asymmetry in baryons.\footnote{Note that this holds after the annihilation of pions at $T\sim 50$ MeV. As will be discussed in great detail in sec. \ref{sec:Update_funcQCD}, at earlier times the lepton asymmetry is stored both in charged as well as neutral leptons.}  
Therefore, a large lepton asymmetry --if existent-- would be stored exclusively in the electrically neutral neutrinos, i.e.
\begin{equation}
    Y_{L_{\alpha}}\simeq \frac{n_{\nu_\alpha}}{s} \,.
\end{equation}

Due to the extremely low energies of the relic neutrinos, direct detection of the cosmic neutrino background is not feasible at the moment and there are no direct constraints on the lepton flavor asymmetries $Y_{L_\alpha}$. 
Nevertheless, they are subject to indirect constraints from cosmological observations which we elaborate on in the following. 

Lepton flavor asymmetries can in general impact the formation of primordial element abundances during Big Bang nucleosynthesis (BBN) and the formation of the cosmic microwave background (CMB) in two distinctive ways (e.g.~\cite{Lesgourgues:1999wu, Beaudet:1976}): Firstly, the Hubble expansion rate is enhanced in the presence of lepton flavor asymmetries. This is usually parameterized in terms of an increase in the effective number of relativistic degrees of freedom $N_{\mathrm{eff}}$ and has an impact on the neutron-to-proton ratio at the onset of BBN. Secondly, the weak interactions that determine the neutron-to-proton ratio also depend on the phase-space distribution of the electron neutrino \cite{Beaudet:1976}. Note that the first effect is neither sensitive to the sign nor the flavor of the lepton asymmetries, whereas the second effect exclusively depends on the electron neutrino asymmetry and its sign. Finally -- while both effects impact the outcome of BBN -- it is the second effect that dominates the production of $^4$He. In other words, the $^4$He abundance is primarily sensitive to the electron lepton asymmetry.

How exactly the observations of $^4$He and $N_{\mathrm{eff}}$ translate into bounds on the lepton flavor asymmetries however strongly depends on the details of neutrino oscillations that become efficient around $T \sim 10$ MeV.

It has first been shown in \cite{Dolgov:2002ab,Wong:2002fa} that neutrino oscillations tend to equilibrate the different flavor asymmetries: While the individual $Y_{L_{\alpha}}$ can be initially very different, they approach a common value during neutrino oscillations (thereby however conserving the total lepton asymmetry $Y_L$). As elaborated in \cite{Pastor:2008ti,Mangano:2010ei,Mangano:2011ip}, if neutrino oscillations start sufficiently early before neutrino decoupling at $T\sim 1$ MeV, the equilibration is almost perfect such that the final lepton flavor asymmetries are $Y_{L_{\alpha}} \approx \frac{Y_L}{3}$. Furthermore, in this case, the final neutrino momentum distributions are thermal, such that $Y_{L_{\alpha}}$ and $N_{\text{eff}}$ can be expressed in terms of neutrino chemical potentials $\mu_{\nu_{\alpha}}$ (or more specifically $\xi_{\nu_{\alpha}}= \mu_{\nu_\alpha}/T_{\nu}$ with $T_{\nu}$ being the neutrino temperature), see e.g. \cite{Mangano:2011ip,Oldengott:2017tzj,Pitrou:2018cgg} for the explicit relations. Under this assumption, the CMB analysis \cite{Oldengott:2017tzj}\footnote{The BBN analysis \cite{Pitrou:2018cgg} as well assumes thermal neutrino spectra and $Y_{\nu_{\alpha}}=\frac{Y_L}{3}$.} found $\xi_{\nu} = -0.002^{+0.114}
_{-0.111}$ (95\%~CL) or
\begin{equation}
 |Y^0_\nu| < 1.2 \times 10^{-2}  \, \Rightarrow  |Y^0_{\nu_{e,\mu,\tau}}| < 4.0 \times 10^{-3}  \hspace{1cm} \text{(perfect equilibration)}. 
    \label{eq:CMB_bound_equil}
\end{equation}
As a large lepton asymmetry can be only stored in the electrically neutral neutrino sector due to the charge and isospin neutrality of the Universe, we can apply the above limit directly for our large lepton asymmetry and take $Y_L \approx Y_\nu$.
It is crucial to note that the bounds on the individual $Y_{L_{e,\mu,\tau}}$ in Eq.~\eqref{eq:CMB_bound_equil} are understood to apply only on their final values after the onset of neutrino oscillations. BBN and the CMB are hence only sensitive to the total lepton asymmetry $Y_L$ and in principle the individual \textit{initial} lepton flavor asymmetries $Y^{ini}_{L_{\alpha}}$ can be much larger as long as their sum fulfills the bound on $|Y_L|$ in Eq. \eqref{eq:CMB_bound_equil}.  

Conceptually more difficult are scenarios, where equilibration is not perfect. As shown in \cite{Pastor:2008ti,Mangano:2010ei,Mangano:2011ip,Barenboim:2016shh,Johns:2016enc,Froustey:2021azz}, depending on the mixing angles (specifically the value of $\sin^2 \theta_{13}$) as well as the mass hierarchy and also the initial asymmetries, neutrino oscillations are potentially happening too close to the time of neutrino decoupling such that equilibration is only partial and the neutrino distribution functions are not thermal any longer. In this case, the lepton flavor asymmetries cannot simply be expressed in terms of chemical potentials any longer and the analyses of \cite{Oldengott:2017tzj,Pitrou:2018cgg} are strictly speaking not valid. The latest analysis of \cite{Froustey:2021azz,Froustey:2024mgf} (taking into account the full 3-flavor description and full mean-field and collision effects) however seems to imply that the approximation of full equilibration is justified, given the precision of nowadays BBN and CMB data.

Interestingly, very recently the EMPRESS survey \cite{Matsumoto:2022tlr} reported a measurement of the $^4$He abundance which is lower by roughly 1 $\sigma$ compared to previous measurements. As reported in \cite{Escudero:2022okz,Matsumoto:2022tlr,NANOGrav:2023hde,Burns:2022hkq}, this could be interpreted as a hint towards a positive asymmetry in the electron neutrino sector. However, since the difference between the new EMPRESS measurement and previous measurements is mainly driven by the new observation of only 5 extremely metal-poor galaxies (EMPG), we wait until future observations of EMPGs shed light on the current situation.

\subsection{Hypermagnetic fields}
In~\cite{Domcke:2022uue} it has been noted that a large lepton flavor asymmetry above a temperature of $10^6$~GeV can lead to chiral plasma instability, sourcing helical hypermagnetic fields. If such helical hypermagnetic fields survive until the electroweak symmetry breaks, then it can result in a baryon number overproduction. This leads to a constraint $|\xi_{\Delta \nu_\alpha}|<0.004$, which is even tighter than the BBN and CMB constraints. However, such a limit is not strictly applicable in the presence of additional symmetries such as a $(\mu+\tau)$-symmetry or if the asymmetry is generated below $10^6$ GeV, therefore should be applied according to the injection time of the lepton asymmetry. We further note that the derivation of this constraint requires a restoration of the electroweak symmetry at high temperatures~\cite{Joyce:1997uy}, which is not the case we are interested in. Therefore, this constraint is not applicable in our scenario.


\section{Symmetry non-restoration and sphaleron freeze-in}
\label{sec:Symmetry non-restoration and sphaleron freeze-in}
The presence of a large (sizeable as compared to the entropy) background charge in the Universe can lead to the non-restoration of the electroweak symmetry at high temperatures. This can be understood from the fact that the Higgs boson condensate, responsible for the spontaneous breaking of the electroweak symmetry, not only depends on the temperature but also on the non-vanishing chemical potential of any fermionic species or in other words on the fermionic charge asymmetry. In this context, one of the main scenarios of interest is a large lepton charge asymmetry while maintaining the electric charge neutrality of the Universe~\cite{Linde:1976kh}. A large lepton asymmetry can lead to the non-restauration of the electroweak symmetry, which leads in turn to an exponentially suppressed rate for sphalerons. As a consequence, the conversion of lepton asymmetry into baryon asymmetry happens extremely slowly, which is crucial in avoiding the overproduction of the baryon asymmetry. We will refer to this mechanism as ``sphaleron freeze-in".

\subsection{The perturbative approach and motivation to go beyond}
The impact of the finite lepton asymmetry on the thermodynamics of the electroweak theory has been discussed in~\cite{Linde:1976kh,Linde:1979pr,Kugo:1982cu,Ferrer:1987jc,Kapusta:1990qc,Khlebnikov:1996vj,Laine:1999wv} using perturbative loop calculations~\cite{Anderson:1991zb,Carrington:1991hz,Dine:1992wr,Arnold:1992rz,Fodor:1994bs} for the effective potential showing that the critical temperature of the electroweak phase transition increases with increasing chemical potential. Large lepton asymmetry as an explanation of the absence of topological defects has been discussed in~\cite{Bajc:1997ky,McDonald:1999he,Bajc:1999he}, while in~\cite{McDonald:1999in,March-Russell:1999hpw,Barenboim:2017dfq} a large lepton asymmetry has been used to explain the observed baryon asymmetry of the Universe. These studies employ the perturbative effective potential at high temperature and finite chemical potentials of baryons and leptons given by~\cite{Bajc:1997ky,Barenboim:2017dfq}
\begin{eqnarray} \label{veff_per}
V_{\rm eff} &=& \frac{\lambda }{4} v^4 + \frac{\lambda'}{2}T^2 v^2 + \frac{g^2}{8} C^2 v^2 + \frac{n_L^2}{T^2}
 + \frac{4 n_L^2 \left( 3 v^2 + 12 C^2 + 14 T^2 \right)}{54 C^2 v^2 + \left( 87 v^2 + 96 C^2 \right) T^2 + 112 T^4}\,,
\end{eqnarray}
where $n_L$ denotes the total lepton asymmetry. Here, $v$ is the vacuum expectation value for the CP-even neutral component of the SM Higgs field and $C\equiv \langle A^1_1\rangle$, with $A^a_\mu$ denoting the gauge potential corresponding to $SU(2)_L$. The coupling $\lambda'$ is given by
\begin{eqnarray}
\lambda' = \frac{1}{12} \left[ 6 \lambda + y_\tau^2 + 3 y_t^2 + 3 y_b^2 + \frac{3}{4} \left( g'^2 + 3 g^2 \right) \right]\,,
\end{eqnarray}
with $\lambda$ indicating the tree-level quartic coupling, $y_f$ the Yukawa couplings of the corresponding SM fermion $f$, $g$ the gauge couplings of $SU(2)_L$ and $g'$ of $U(1)_Y$. From Eq.~\eqref{veff_per} it is straightforward to see that $V_{\rm eff}$ can be minimized as a function of $v(T)$ and $C(T)$ for given values of the lepton number density asymmetry $n_L$, see e.g.~\cite{Barenboim:2017dfq}. From such an estimate it is apparent that for the non-restoration of the electroweak symmetry, an asymmetry of order $Y_L=n_L/s\sim 1$ is required. However, such a large lepton asymmetry is subject to several stringent constraints as we have discussed already (see Eq.~\ref{eq:CMB_bound_equil}). 

While the perturbative calculations discussed above do provide a qualitative idea, such an approach is inconsistent due to the infrared divergences arising while integrating over the bosonic zero (static) modes in the path integrals~\cite{Linde:1980ts}.  A full prescription to deal with the infrared divergences at finite temperature was presented in~\cite{Kajantie:1995dw}, by matching the full 4-dimensional theory to an effective 3-dimensional theory, which can then be solved with lattice Monte Carlo methods~\cite{Kajantie:1995kf,Kajantie:1996mn,Kajantie:1996qd,Karsch:1996yh,Gurtler:1997hr}. This approach was further generalized to finite chemical potential in~\cite{Gynther:2003za}. In what follows, we will adopt the prescription of~\cite{Gynther:2003za} to compute the effective potential at finite temperature and finite chemical potentials which can be minimized numerically to solve for $v(T)/T$ as a function of the lepton asymmetry. We will then combine these numerical results with an exact computation of the small-fluctuation determinant~\cite{Carson:1990jm} to estimate the sphaleron freeze-in of the baryon asymmetry and to study the viability of a large lepton asymmetry as an explanation of the observed baryon asymmetry. We find that this approach significantly reduces the amount of the primordial asymmetry required for a successful baryogenesis scenario as compared to the previous predictions using the perturbative approach~\cite{Gao:2023djs}.


\subsection{Standard Model effective potential at finite lepton asymmetry using dimensional reduction}
Several methods exist to resum the theory such that the hard modes screen the infrared modes. One convenient method that automatically organizes perturbation theory is provided by dimensional reduction~\cite{Kajantie:1995dw}. In the reduction of the full 4-dimensional theory to a 3-dimensional one, all the fermionic and bosonic modes with $n\neq 0$ ($n$ being the index of Fourier expansion of the fields into Matsubara modes) acquire a mass $m\sim \pi T$ and they are referred to as superheavy modes. The temporal components of the gauge bosons acquire a mass $m \sim gT$ and are referred to as heavy modes. The remaining fields with mass $m\sim (g^2T)$ are referred to as light modes~\cite{Kajantie:1995dw}.  At high temperatures the perturbative expansion parameter $\frac{g^2 T}{E}$ becomes large for such light modes. This is in contrast to the other heavier modes (non-static modes with mass $m\sim \pi T$ or static modes with $|p|>gT$) leading to a consistently small expansion parameter.
In the dimensionally reduced theory, heavy Matsubara modes are integrated out to define an effective potential at a soft scale given by $gT$. The temporal component of the vector bosons acts as scalar fields with mass $O(gT)$ at this scale and themselves can be integrated out to give an effective potential at an ultrasoft scale $O(g^2 T/\pi)$. Resummation is included by construction due to the matching master formula~\cite{Kajantie:1995dw} 
relating four-dimensional fields $\varphi$ to their three-dimensional counterparts $\varphi_3$ at finite temperatures,
\begin{equation}
    \varphi ^2_{3} = \frac{1}{T} \varphi ^2 [1+\bar{\Pi} ^\prime (0)] \, ,
\end{equation}
where $\Pi$ is the self-energy, and the prime denotes the derivative of the self-energy w.r.t. external momentum.
The renormalized 2-point Green's function for a light scalar field in full 4d theory of the form
\begin{equation}
k^2+m_S^2+\Pi(k^2)=k^2+m_S^2+\Pi_3(k^2)+\bar{\Pi}(k^2) \, ,
\end{equation}
is matched to the corresponding three-dimensional theory equivalent
\begin{equation}
k^2+m_3^2+\Pi_3(k^2) \, ,
\end{equation}
with $\Pi_3(k^2)$ denoting the self-energy contribution due to light and the heavy modes only and $\bar{\Pi}(k^2)$ denoting all other contributions involving super-heavy degrees of freedom. Since the super heavy modes can be integrated out without any IR problems, $\bar{\Pi}(k^2)$ can be expanded in the external momenta in the limit $k\ll T$ as
\begin{equation}
\bar{\Pi}(k^2)=\bar{\Pi}(0)+\bar{\Pi}^{'}(0)\, k^2 +{\mathcal{O}}(g^2\frac{k^4}{T^2})\, .
\end{equation}
The corresponding masses are related up to order $g^4$ via
\begin{equation}
m_3^2=\left[ m_S^2 +\bar{\Pi}(0) \right] \left[ 1- \bar{\Pi}^{'}(0)\right]\, .
\end{equation}
 We note that the 3d theory is by construction super-renormalizable, and the relationships between the 3d and 4d parameters are determined appropriately. For instance, the 4d couplings and masses are dependent on the 4d renormalization scale. However, the 3d couplings are related to the full 4d theory such that they are renormalization scale independent, while the scalar masses can have mass divergences, which can be renormalized by a 3d renormalization scale independent of the 4d renormalization scale. This is because the bare mass parameters produced via the dimensional reduction are 4d theory renormalization scale independent.

At finite density, however, the field renormalization that appears in the matching conditions of the two-point Green's functions between 4d and 3d theories 
become functions of the chemical potentials. This results in chemical potential dependent corrections to the finite temperature $\mu=0$ masses and couplings. Secondly, additional terms are induced due to the reduction of symmetries in the four-dimensional theories as compared to the $\mu=0$ finite temperature case. In particular, in the presence of non-vanishing chemical potentials, there are $C$ violating (but $P$ and $T$ conserving) terms in the path integral which lead to additional terms in the dimensionally reduced effective field theory.

The electroweak theory at finite temperature can be expressed by the Euclidean action
\begin{eqnarray}{\label{eq:action4d}}
S & = & \int_{0}^{\beta} d\tau \int d^3x \, {\cal L} \,
\end{eqnarray}
defined on a finite time interval $0<\tau<\beta \equiv\frac{1}{T}$, where
\begin{eqnarray}{\label{eq:lag4d}}
{\cal L}  & = &(D_\mu\Phi)^\dagger D_\mu\Phi - \nu^2\Phi^\dagger \Phi + \lambda(\Phi^\dagger \Phi)^2
+ \frac{1}{4}G_{\mu\nu}^aG_{\mu\nu}^a  + \frac{1}{4}F_{\mu\nu}F_{\mu\nu}
+ \bar{{l}}_L\slashed{D}{l}_L + \bar{e}_R\slashed{D}e_R  \\
&+&  \bar{{q}}_L\slashed{D}{q}_L + \bar{u}_R\slashed{D}u_R + \bar{d}_R\slashed{D}d_R + g_Y\left(\bar{{q}}_L\tilde{\Phi}t_R 
+ \bar{t}_R\tilde{\Phi}^\dagger{q}_L \right). 
\end{eqnarray}
Here ${q}_L$ and ${l}_L$ denote the $SU(2)_L$ doublet quark and lepton fields. $u_R$, $d_R$ and $e_R$ denote the $SU(2)_L$ singlet right-handed quark and lepton fields. $\Phi$ denotes the SM Higgs doublet with $\tilde{\Phi} = i\tau^2\Phi^\ast$. The covariant derivative is defined as $D_\mu \equiv \partial_\mu + IigA_\mu^a\tau^a + Yig'B_\mu$, with $I$ and $Y$ denoting the isospin and hypercharge, respectively. $g_Y$ is assumed to be nonzero only for the top quark. The field strength tensors are defined as $G^a_{\mu\nu} \equiv \partial_\mu A_\nu^a - \partial_\nu A_\mu^a 
- g\epsilon^{abc}A_\mu^b A_\nu^c$, and $F_{\mu\nu} \equiv \partial_\mu B_\nu - \partial_\nu B_\mu$, with  $A_\mu^a$ and $B_\mu$ denoting the gauge fields corresponding to $SU(2)_L$ and $U(1)_Y$. Considering all the conserved currents, the relevant partition function can be expressed as~\cite{Kapusta:1990qc}
\begin{eqnarray}
{\cal Z} & = & \text{Tr} \,\exp\big(-\beta(H-\mu_i X_i-\mu_Y Q_Y - \mu_{T^3}Q_{T^3})\big) \label{eq:Z1}\\
& \equiv &\int\!\!{\cal D}\varphi\exp\left[-S + \int_0^\beta d\tau\;\left(\mu_B B + \sum_{i=1}^{n_f}\mu_{L_i} L_i\right)\right] \label{eq:Z2}
\end{eqnarray}
where $\varphi$ corresponds to the set of all the fields and
\begin{eqnarray}
S & = & S\left[\tilde{\varphi},B_0+\frac{i\mu_Y}{g'},A_0^3+\frac{i\mu_{T^3}}{g}\right]\\
\mu_B & \equiv & \frac{1}{n_f}\sum_{i=1}^{n_f}\mu_i, \quad  \mu_{L_i} \equiv -\mu_i,   
\end{eqnarray}
where $\tilde{\varphi}$ is defined such that it excludes $B_0$ and $A_0^3$ from the set $\varphi$. The globally conserved currents corresponding to $n_f$ flavors are defined as
\begin{equation}
X_i = \frac{1}{n_f}B - L_i, \quad i=1\dots n_f
\end{equation}
with the baryon and lepton number currents defined as
\begin{eqnarray}
B & = & \frac{1}{3}\sum_{f,c}\int d^3x \bar{q}_{c,f}\gamma_0 q_{c,f} \, ; \quad \quad 
L_i  =  \int d^3x \left(\bar{e}_i\gamma_0 e_i + \bar{\nu}_i\gamma_0 P_L\nu_i\right) \, ,
\end{eqnarray}
  where $P_L$ denotes the usual left-handed projection operator. 

A $3$-dimensional $\mathrm{SU}(2)\times \mathrm{U}(1)$ invariant effective theory involving a fundamental scalar doublet (Higgs) and four adjoint scalars (corresponding to the temporal components of the gauge fields in the fundamental $4$-dimensional theory) can first be obtained by integrating out the nonzero Matsubara modes.  The general form for such an effective theory is given by the Lagrangian~\cite{Gynther:2003za}
\begin{eqnarray}
{\cal L}_1 & = & \frac{1}{4}G_{ij}^a G_{ij}^a + \frac{1}{4}F_{ij}F_{ij} + \left(D_i\Phi\right)^\dagger D_i\Phi 
+ m_3^2\Phi^\dagger\Phi + \lambda_3\left(\Phi^\dagger\Phi\right)^2 
+\frac{1}{2}\left(D_i A_0^a\right)^2  + \frac{1}{2}m_D^2A_0^aA_0^a \nonumber \\
&+ & \frac{1}{4}\lambda_A A_0^aA_0^aA_0^bA_0^b + \frac{1}{2}\left(\partial_i B_0\right)^2 + \frac{1}{2}m_D^{\prime 2}B_0^2 
+ h_3 \Phi^\dagger\Phi A_0^aA_0^a + h_3'\Phi^\dagger\Phi B_0^2 + \frac{1}{2}g_3 g_3' B_0 \Phi^\dagger A_0^a\tau^a\Phi \nonumber\\
&+& \alpha \epsilon_{ijk}\left(A_i^aG_{jk}^a - \frac{i}{3}g_3\epsilon^{abc}A_i^a A_j^b A_k^c\right)  + \alpha'\epsilon_{ijk}B_i F_{jk} + \kappa_1 B_0 + \rho \Phi^\dagger A_0^a\tau^a\Phi 
+ \rho' \Phi^\dagger\Phi B_0 \nonumber\\ &+& \rho_GB_0A_0^aA_0^a, \nonumber\\
\label{eq:1eft3d}
\end{eqnarray}
where the terms in the last two lines of Eq.~\eqref{eq:1eft3d} correspond to new $CP$ and $CPT$ violating contributions in the presence of finite chemical potential, $D_iA_0^a = \partial_iA_0^a-g_3\epsilon^{abc}A_i^bA_0^c$. The relevant effective potential for the Higgs expectation value $\varphi$ up to order ${\cal O}(\varphi^4)$ is then given by~\cite{Gynther:2003za}
\begin{eqnarray}
V_\mathrm{eff}(\varphi) & = & \frac{1}{2}\left(m_3^2 -\frac{1}{2\pi}h_3m_D + \frac{h_3'\kappa_1^2}{m_D^{\prime 4}}\right)\varphi^2  + \frac{1}{4}\left(\lambda_3 -\frac{1}{4\pi}\frac{h_3^2}{m_D} 
- \frac{2h_3'\kappa_1^2}{m_D^{\prime 4}} \left(\frac{h_3}{m_D^2}+\frac{h_3'}{m_D^{\prime 2}}\right)\right)\varphi^4 \nonumber \\
& & - \frac{1}{32\pi}\left(2g_3^3+(g_3^2+g_3^{\prime 2})^{3/2}\right)\varphi^3 \, .
\label{eq:feft3d}
\end{eqnarray}
To obtain Eq.~\eqref{eq:feft3d}, the contributions from the terms in Eq.~\eqref{eq:1eft3d} with the couplings
$\lambda_A$ and $\rho$ are neglected since they follow the power counting rules $\lambda_A\sim g^4$ and $\rho_G\sim g^{7/2}$, respectively, and therefore are small. The terms in Eq.~\eqref{eq:1eft3d} with the couplings $\rho'$  and $\rho_G$ are also neglected since the three-point vertices are negligible as compared to similar vertices induced by the four-point vertices $\Phi^\dagger\Phi B_0^2$ and $B_0\Phi^\dagger A_0^a\tau^a\Phi$ in combination with the term $\kappa_1 B_0$. The corrected Debye masses $m_D$, $m'_D$, coupling constants $\lambda_3$, $h_3$, $g_3$ and scalar mass parameter $m_3^2$ as well as the coefficient $\kappa_1$ of the new term in the effective theory due to the presence of a finite chemical potential can be found in~\cite{Gynther:2003za}, while the expressions for the above quantities as well as for $h'_3={g'}_3^2/4$ for zero chemical potential (note that the self-coupling of the adjoint scalar does not receive any corrections due to the chemical potential) are given in~\cite{Kajantie:1995dw}. In Appendix \ref{app:effpot} we reiterate the full expressions that we use for our computations for easy reference.

\begin{figure}
    \centering
\includegraphics[width=0.49\textwidth]{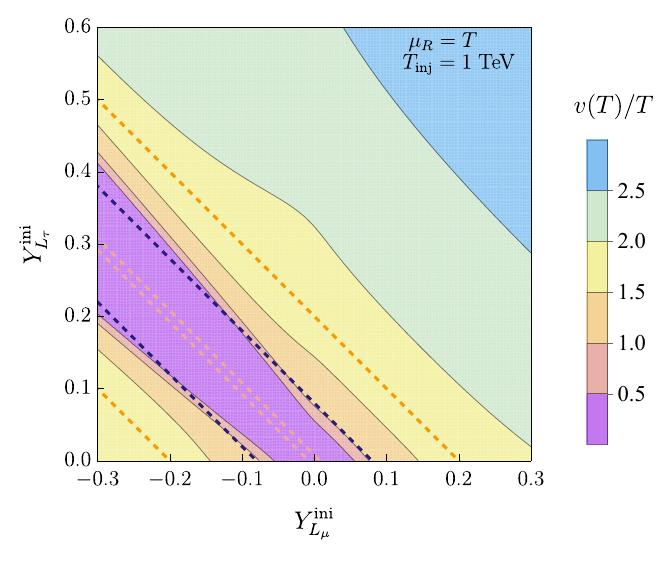}
    \includegraphics[width=0.49 \textwidth]{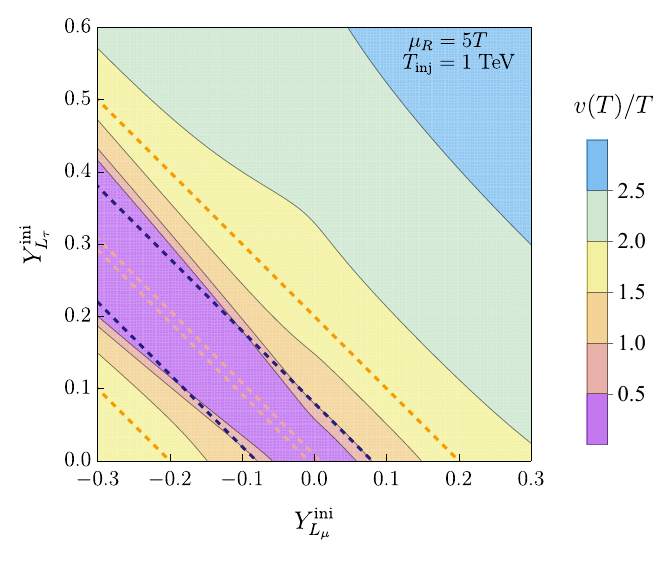}
    \includegraphics[width=0.49\textwidth]{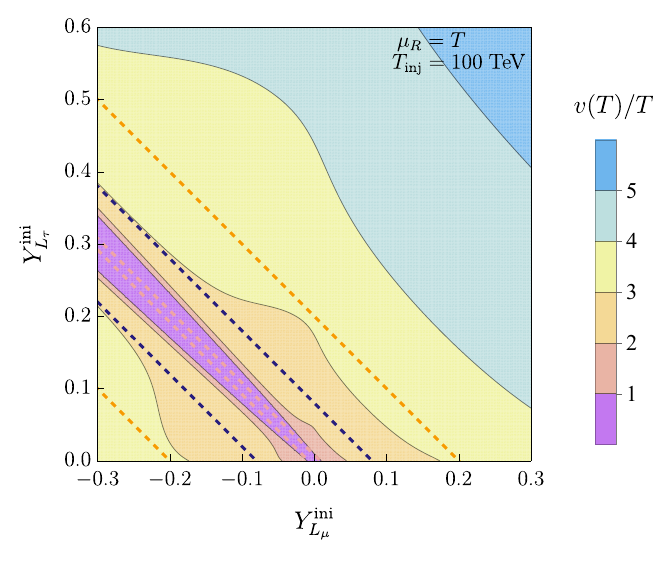}
    \includegraphics[width=0.49 \textwidth]{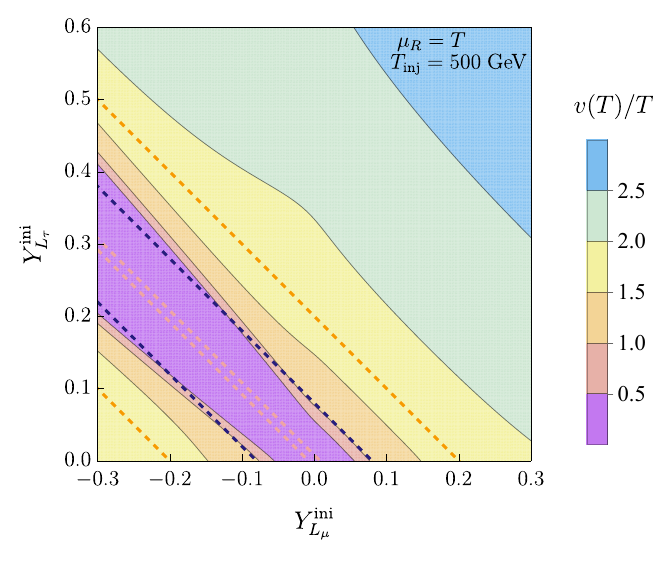}
    \caption{$v(T)/T$ contours in the plane of $Y_{\nu_\mu}$ and $Y_{\nu_\tau}$ obtained by minimizing the dimensionally reduced effective potential at finite temperature and chemical potential. To demonstrate the dependence of the result on the choice of regularization scale, we vary the regularization scale by a factor of five, showing the cases $\mu_R=T$ (top-left) and $\mu_R=5T$ (top-right) for $T_\text{inj}=1$~TeV. To demonstrate the dependence on the injection temperature, we show the cases $T_\text{inj}=100$~TeV (bottom-left) and $T_\text{inj}=500$~GeV (bottom right) with $\mu_R=T$. The region between the pink, blue, and orange pair of dashed lines indicates the allowed parameter space based on CMB and BBN constraints for the cases $\Delta=1$ (no entropy dilution), $\Delta=10$, and $\Delta=25$, respectively. See sections~\ref{sec:freeze-in_details},~\ref{sec:Late time entropy dilution:} and appendix~\ref{sec:Entropy dilution} for discussion on entropy dilution. }
    \label{fig:pot_min}
\end{figure}

We use the package SARAH~\cite{Staub:2008uz} to calculate the evolution of running couplings matched at the Z-pole \footnote{ Note that our conventions slightly differ from the original SARAH~\cite{Staub:2008uz} convention in that the Higgs quartic is rescaled by a factor of 2 and the U(1) hypercharge gauge coupling constant is rescaled by a factor of $\sqrt{5/3}$.}. The RGEs we use are only functions of the numerically most important couplings ${\lambda, \mu ^2 , g_i, y_t}$ whose values at the Z-pole are obtained from~\cite{Huang:2020hdv} (employing multiloop matching) and are tabulated in Appendix~\ref{app:matching}. Special attention is also given to the top mass as described in Appendix~\ref{app:matching}. This is due to the slow convergence of perturbation theory -- both the Yukawa coupling and the strong couplings are $\mathcal{O}(1)$. We then minimize the 3d effective potential in Eq.~\eqref{eq:feft3d} numerically. Given the tight constraints from the CMB and BBN, the maximal allowed $Y_{\nu_e}$ will have neglible impact on our analysis such that we make the simplifying assumption of a vanishing $Y_{\nu_e}$, without any loss of generality.  Using this approach, we first find a numerical interpolation of $v(T)/T$ contours in the plane of $Y_{\nu_\mu}$ and $Y_{\nu_\tau}$. In Fig.~\ref{fig:pot_min}, we show a contour plot showing $v(T)/T$ corresponding to the minimized potential for the lepton asymmetry injection scale $T=1$~TeV as a function of $Y_{\nu_\mu}$ and $Y_{\nu_\tau}$, for two choices of regularization scale $\mu_R$, showing that our results are fairly regularization scale independent.
Note that the potential we employ neglects two-loop effects. For further discussion regarding their impact we refer to Refs.~\cite{Croon:2020cgk,Gould:2021oba}.

\subsection{Sphaleron freeze-in baryogenesis and generation of a large lepton asymmetry}\label{sec:freeze-in_details}

To obtain an accurate estimate of the baryon asymmetry from a large lepton asymmetry, we combine our obtained $\frac{v(T)}{T}$ by numerically minimizing the finite temperature potential with an exact computation of the determinant of small fluctuations around the sphaleron configuration of electroweak theory in the high temperature limit (denoted by $\kappa$, see below for definition), see e.g.~\cite{Carson:1990jm}, to estimate the sphaleron freeze-in rate. The sphaleron rate per unit of time is given by~\cite{Carson:1990jm}
\begin{equation}
    \Gamma_{\rm sph} =\frac{\omega _-}{2 \pi T^3} \left( \frac{v^2}{T} \right)^3 {\cal N} _{\rm rot} {\cal V}_{\rm rot} {\cal N}_{\rm tr} \kappa e^{-\frac{4 \pi v }{g_2T} E_{\rm sph}} \ . 
\end{equation}
Here $\omega_{-}$ is the negative-mode frequency dictating the rate of decay in small fluctuations around the sphaleron determined in the units of $gv$ as a function of $\lambda/g^2$ . $E_{\text{sph}}$ is an $\mathcal{O}(1)$ number dependent on $\lambda/g^2$, determining the energy of the sphaleron in the units of $4\pi v/g$. The quantities ${\cal N}_{\rm tr}$ and ${\cal N}_{\rm rot}$ are normalization integrals relating the natural coordinates describing the translations and rotations of the sphaleron to canonical coordinates. The volume of rotation group is ${\cal V}_{\rm rot}=8 \pi ^2$, and $\kappa$, the determinant of small fluctuations, is given by
\begin{equation}
    \kappa = {\rm Im} \left[ \frac{{\rm det}[\delta ^2 S_{gf}/\delta \phi ^2]_{\phi = \phi _{\rm vac}} \Delta _{\rm FP,\phi=\phi_{\rm sp}}}{{\rm det}[\delta ^2 S_{gf}/\delta \phi ^2]_{\phi = \phi _{\rm sp}} \Delta _{\rm FP,\phi=\phi_{\rm vac}}} \right]\,,
\end{equation}
which contains the Faeddeev-Popov determinant, $\Delta_{\text{FP},x}$, where $x \in \{ \rm sp, vac\}$ are the sphaleron and vacuum field configurations, respectively. 

Physically, $\kappa$, measures the number of non-zero frequency modes available in the vicinity of the sphaleron relative to that of the perturbative vacuum. For the standard model these quantities, as well as $({\cal N}_{\rm tr},{\cal N}_{\rm rot})$, and the negative frequency mode, $\omega _-$ have been calculated as a function of $\lambda / g^2$~\cite{Carson:1990jm}.  Similarly, we take the sphaleron energy as a function of the same ratio from~\cite{Klinkhamer:1984di}. The final baryon asymmetry at $T=T_{\rm fin}$ generated via the sphaleron freeze-in can then be estimated as
\begin{equation}
  Y_B^{\text{fin}}\equiv Y_B (T_\text{fin})\simeq - \big(2 H(T_\text{fin})\big)^\frac{3}{2}\int^{T_\text{fin}}_{T_\text{inj}} dT \frac{1}{\big(2 H(T)\big)^\frac{3}{2}}\frac{1}{2 H(T) T} \;\Gamma_{\rm sph}\big(v(T),T\big)\; Y_L^{\rm ini}\; ,
\end{equation}
where $H(T)$ denotes the Hubble rate and an initial lepton asymmetry $Y_L^{\rm ini}$ is assumed to be injected instantaneously into the thermal bath at $T_{\rm inj}$, leading to a non-restoration of the electroweak symmetry.  Utilizing the above prescription, we obtain the generated baryon asymmetry as a function of the initial lepton asymmetry distributed into lepton flavor asymmetries in $\mu$- and $\tau$-flavors ($Y_{L_{\mu (\tau)}}^{\text {ini}}$). Note that in the presence of flavor violating processes among $\mu$- and $\tau$-flavors there can further be corrections to the individual flavour asymmetries, which we assume to be negligible in our analysis~\footnote{In addition, in the full chemical potential equilibration limit the charged lepton asymmetry will lead to a Higgs asymmetry, dictated by the total electric charge and third component of the isospin conservation of the universe. Such a Higgs asymmetry can induce secondary flavour asymmetries in leptons through the Yukawa couplings in equilibrium. However, since the lepton asymmetries are dominantly stored in electric charge-neutral neutrinos (compared to charged leptons), we find the effect to be secondary and negligible. See also section~\ref{sec:Cosmic trajectory and state of the art} for some discussion on the chemical potential relations.}

\begin{figure}
    \centering
    \includegraphics[width=0.49\textwidth]{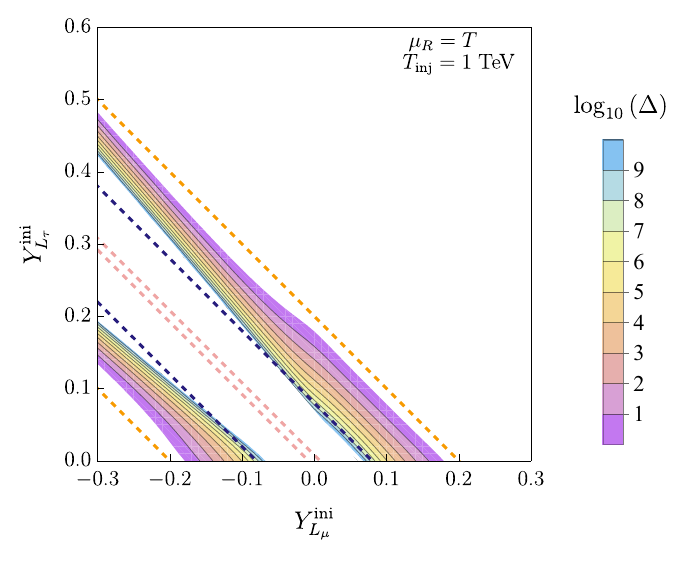}
    \includegraphics[width=0.49\textwidth]{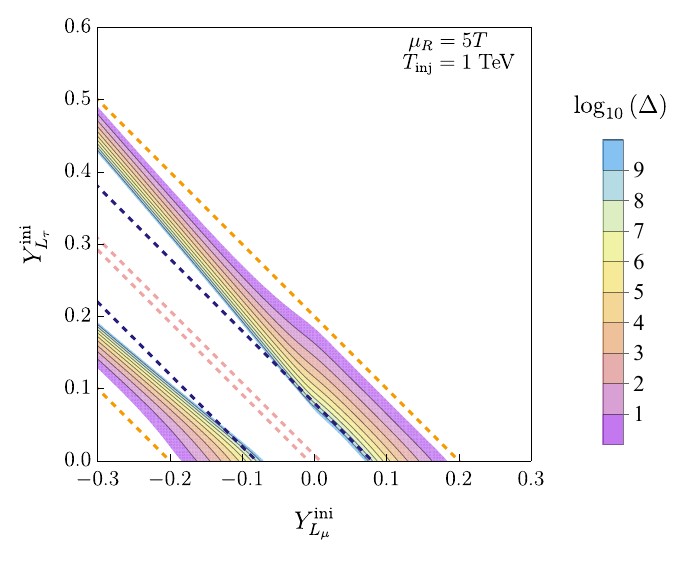}
      \includegraphics[width=0.49\textwidth]{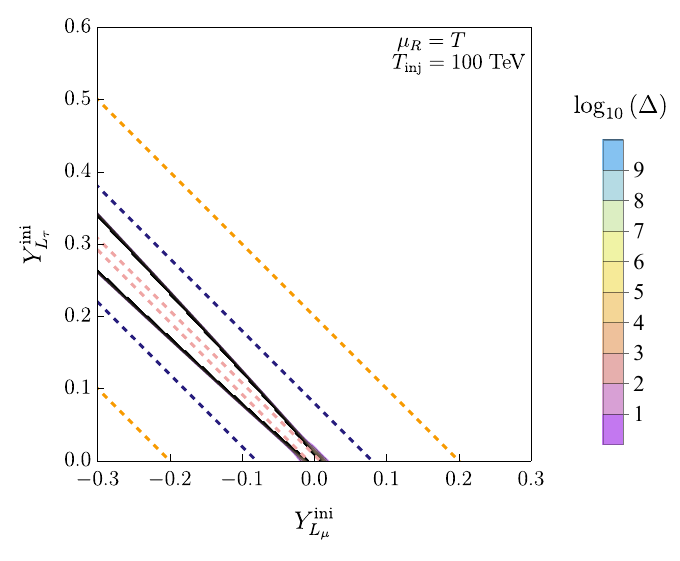}
    \includegraphics[width=0.49\textwidth]{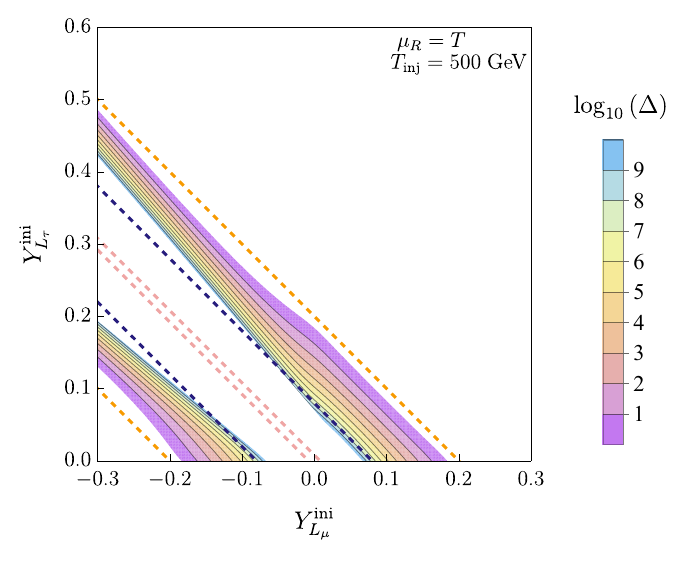}
    \caption{The viable parameter space for successful baryogenesis via sphaleron freeze-in in the plane of $Y_{\nu_\mu}$ and $Y_{\nu_\tau}$ obtained using the dimensionally reduced effective potential at finite temperature and chemical potential (c.f. Fig.\ref{fig:pot_min}) and employing the sphaleron rate computation using the fluctuation determinant. The different colored contours as indicated in the plot legend show the regions corresponding to different values of entropy dilution required to produce the correct observed baryon abundance today. The region between the pink, blue, and orange pair of dashed lines indicates the allowed parameter space based on CMB and BBN constraints for the cases $\Delta=1$ (no entropy dilution), $\Delta=10$, and $\Delta=25$, respectively. See the text and also section~\ref{sec:Late time entropy dilution:} and appendix~\ref{sec:Entropy dilution} for more discussion on entropy dilution. To demonstrate the dependence of the result on the choice of regularization scale, we show the cases $\mu=T$ on the top-left and $\mu=5T$ on the top-right, for the lepton asymmetry injection temperature $T_\text{inj}=1$~TeV. To show the dependence on the injection temperature in bottom-left and bottom right we show the cases $T_\text{inj}=100$~TeV and $T_\text{inj}=500$~GeV, respectively, setting $\mu=T$. We note that in the bottom-left plot the two dark bands result from the shrinking of the colored bands shown in the plot legend.} 
    \label{fig:pot_genesis}
\end{figure}

In Fig.~\ref{fig:pot_genesis}, we plot the contours for $Y_B^{\text{fin}}/Y_B^{\rm obs}$ in the $Y_{L_{\mu (\tau)}}^{\text {ini}}$ plane, where $Y_B^{\rm obs}$ denotes the observed baryon asymmetry today. For the contours with $Y_B^{\text{fin}}/Y_B^{\rm obs}>1$, an entropy dilution, following the asymmetry generation, defined as $\Delta\equiv Y_{B}^{\text {fin}}/Y_{B}^{\rm obs}=Y_{L}^{\text {fin}}/Y_{L}\simeq Y_{L}^{\text {ini}}/Y_{L}$ (see the following sections and appendix for a more detailed discussion) is needed in order not to overproduce the baryon asymmetry of the Universe today. Further, $Y_{L}$ denotes the lepton asymmetry today. The color-coded contours in Fig.~\ref{fig:pot_genesis} delineate the acceptable region capable of reproducing the correct observed baryon asymmetry and the associated entropy dilution required to not over-produce baryon asymmetry. For instance, the darkest violet region represents the parameter space capable of producing the correct observed baryon asymmetry via sphaleron freeze-in for an entropy dilution of $\Delta<10$ following the electroweak phase transition and before BBN. 

To demonstrate the dependence of the final result on the choice of regularization scale, we show the cases $\mu=T$ on the top-left and $\mu=5T$ on the top-right figures, for a lepton asymmetry injection temperature of $T_\text{inj}=1$~TeV. As can be seen by comparing these figures, the final results are fairly regularization scale independent, thereby showing the robustness and reliability of our dimensional reduction based approach. 

The parameter space for successful baryogenesis is also dependent on the temperature when the large lepton asymmetry is injected into the thermal bath. To show the dependence on the injection temperature in bottom-left and bottom-right we show the cases $T_\text{inj}=100$~TeV and $T_\text{inj}=500$~GeV, respectively. We note that the required total asymmetry (i.e. the absolute value of the sum of the asymmetries in two flavors) and the needed entropy dilution decreases with increasing injection temperature of the initial lepton asymmetry. Furthermore, as the injection temperature becomes higher, the parameter space providing the correct observed baryon asymmetry also becomes much more narrow with the differently colored regions shrinking very tightly together, as can be seen from the bottom-left figure.

Remarkably, our approach predicts a reduction by an order of magnitude in the required primordial lepton asymmetry necessary for successful baryogenesis compared to prior findings in the literature, such as Refs.~\cite{Barenboim:2017dfq,Bajc:1997ky}.

As mentioned already, the primordial asymmetry $Y_{L_{\mu (\tau)}}^{\text {ini}}$ may experience an entropy dilution following the electroweak phase transition and before the BBN in the presence of entropy production as discussed in the following section and appendix \ref{sec:Entropy dilution}. We will refer to this as late-time entropy dilution. We find a minimum entropy dilution of 
$\Delta_{\text{min}}\sim \mathcal{O}(10)$ is required to reproduce the correct baryon symmetry. The needed entropy dilution shows a gradual dependence on the lepton asymmetry injection temperature as can be seen by comparing the bottom-left plot with the other ones Fig.~\ref{fig:pot_genesis}, showing that the required entropy dilution goes down with the increasing lepton asymmetry injection temperature.

Note that the regions between the dashed pink, blue, and orange dashed lines represent the allowed parameter space that is in agreement with CMB and BBN constraints (assuming full equilibration among flavors during neutrino oscillations) for the scenarios of $\Delta=1$ (no entropy dilution), $\Delta=10$, and $\Delta=25$, respectively. A larger entropy dilution means that one can start with a larger lepton asymmetry which gets diluted by a factor of $\Delta$ at some time after the injection of the asymmetry (see section \ref{sec:injection}) and before the CMB and BBN, such that the tight constraint on the individual flavours can be satisfied. As we have discussed in section~\ref{sec:BBNandCMBconstraints}, the neutrino oscillations equilibrate the neutrino flavour asymmetry shortly before the CMB and BBN. The perfect equilibration among the neutrino flavours provides the most realistic and tight constraints because of the stong constraint on the electron neutrino flavor from the BBN.

Naively, an ad-hoc large lepton asymmetry and the late-time entropy dilution might appear unrealistic. However, in the following we will discuss how these two ingredients can naturally occur without affecting the dynamics of the electroweak symmetry non-restoration considered above.

\subsection{Injection of large lepton asymmetry in the thermal bath} \label{sec:injection}
One of the main criteria for realizing the proposed scenario is to not create too large baryon asymmetry (same order as the lepton asymmetry) as the large lepton asymmetry is slowly accumulated over time. This is so, because if the sphalerons are in equilibrium and unsuppressed, then any slow means of accumulation of lepton symmetry (e.g. by conventional thermal leptogenesis) will result in the fast conversion of the lepton asymmetry into baryon asymmetry before enough lepton asymmetry can be generated to freeze-in the sphalerons. One of the most interesting scenarios is then provided by the case where some form of decoupled condensate capable of storing a large lepton number (to decay at a later time to populate the SM thermal bath) is formed. For instance, the large lepton asymmetry can be injected via a heavy species condensate, which is decoupled from the thermal bath initially. The thermalization of the lepton asymmetry must then happen instantaneously after the asymmetry injection, such that the large lepton asymmetry non-restores the electroweak symmetry as soon as it gets injected into the thermal bath, implying that the rate of sphalerons is always suppressed, thereby avoiding the generation of a large baryon asymmetry through unsuppressed sphalerons.

This can be achieved, for instance utilizing the Affleck-Dine mechanism~\cite{Affleck:1984fy}. The minimal ingredients are an inflaton field ($\psi$) and a decoupled scalar field ($S: m_S>>T_{\text{th}}$, where $T_{\text{th}}$ is the thermal bath temperature) which provides a flat direction of the potential during inflation. $S$ can either be capable of creating lepton number via CP-violating decays or alternatively, it can possess effective lepton number violating interactions which can lead to a portal for lepton asymmetry generation from a condensate of $S$ fields. (i) The inflaton starts oscillating (when $H \sim m_\psi$) and makes the Universe matter dominated. (ii) $S$ starts oscillating when $H \sim m_S < m_\psi$).
(iii) The inflaton decays at $t\sim (M_P^2/m_\psi^3)$ making the universe radiation dominated. (iv) The $S$ condensate decays and generates directly a lepton asymmetry or via a lepton number violating portal after the Universe becomes radiation-dominated following the inflaton decay. (v) The lepton asymmetry is thermalized once the Hubble rate after inflation falls below the interaction rate of leptons produced. In supersymmetric theories, sneutrinos could be an example for $S$~\cite{Casas:1997gx}. 

In the above discussion, we assume that the electroweak symmetry breaking is instantaneous upon injecting a large lepton symmetry. However, if the condensate of heavy particle species producing the lepton asymmetry decays perturbatively with a decay width $\Gamma_{\text{con}}$, a time scale of about $\Gamma_{\text{con}}^{-1}$ will be needed to build up sufficient lepton asymmetry before the sphaleron interactions freeze-in. In such a scenario, the baryon asymmetry generated before the quenching of the sphaleron interactions is expected to be relatively large, scaling as $\frac{\Gamma_{\text{sph}}}{\Gamma_{\text{con}}} \left(\frac{4\pi E_{\text{sph}} v}{g_2T}\right)^{-2}$, when normalised to the lepton asymmetry building up following an exponential decay law for condensates. Here $\Gamma_{\text{sph}}$ refers to the sphaleron rate per unit thermal volume in the unbroken phase. One interesting possibility to circumvent such a problem could be to implement a scenario where the heavy condensate (that stores the lepton number before producing the lepton asymmetry at late times) can already break the electroweak symmetry, making the sphalerons already less effective before the lepton asymmetry is produced and thermalised. A condensate in the form of an extra Higgs doublet, which solely interacts with SM neutrinos, producing their Dirac mass, could potentially play this role \footnote{We thank the anonymous referee for bringing this possibility to our attention. }. A detailed exploration of such a scenario is beyond the scope of the current work; however, it would surely be an interesting possibility to explore.

\subsection{Late time entropy dilution}
\label{sec:Late time entropy dilution:}
To produce the correct observed baryon asymmetry via the sphaleron freeze-in mechanism, late-time mild entropy dilution ($\Delta\sim \mathcal{O}(10)$) must occur after the electroweak symmetry breaking and before the onset of neutrino oscillations and the BBN. However, it is important to note here that even in the absence of any entropy dilution the possibility of a first-order QCD phase transition with a potential gravitational wave signal remains a viable possibility (independent of whether a successful baryogenesis scenario can be realized or not).

To realize the entropy dilution, one of the basic possibilities is to employ again a late-time decaying state decoupled from the thermal bath which comes to dominate the energy density of the Universe at a late time (after the asymmetry generation and before the BBN) and then decay to produce a late time reheating leading to entropy dilution. We note that the possibility of entropy dilution mechanism accomplished by a second phase of inflation is also discussed in the literature~\cite{Lyth:1995ka,Davoudiasl:2015vba}. One of the well-discussed possibilities is the saxion associated with a Peccei-Quinn symmetry\cite{Barenboim:2017dfq}. Many other naturally motivated candidates include e.g. long-lived moduli \cite{Moroi:1999zb} (one of the major challenges of such scenarios being rather large entropy release in all species), gravitinos \cite{Moroi:1994rs}, curvatons \cite{Moroi:2002rd}, dilatons \cite{Lahanas:2011tk}, $Q$-balls \cite{Fujii:2002kr}, etc.. In Appendix~\ref{sec:Entropy dilution}  we provide a detailed discussion and the relevant formulae regarding the dynamics of entropy dilution due to a late-time decaying state decoupled from the thermal bath for future model building purposes. 

We want to emphasize that while our scenario does involve interesting and nontrivial cosmological model-building possibilities, the new physics beyond the SM invoked to realize the lepton asymmetry injection and late time entropy dilution can be very naturally decoupled from the SM sector realizing the sphaleron freeze-in mechanism. 

\section{Cosmic QCD epoch}
\label{sec:Cosmic QCD epoch}
The QCD phase diagram charts at which temperature $T$ and chemical potential $\mu$ baryonic matter exists in the form of either free quarks and gluons or in the form of hadrons. Lattice QCD calculations have revealed that at zero chemical potential, the transition between both phases happens smoothly, in terms of a cross-over. Due to the infamous sign problem, lattice QCD is however known to be only exact at vanishing chemical potential, leaving the QCD diagram at finite chemical potential unsettled. Remedy can be provided by continuum QCD methods -- i.e. functional QCD (fQCD) methods including Dyson-Schwinger equations (DSEs) \cite{Binosi:2009qm,Roberts:2000aa,Eichmann:2016yit, Fischer:2018sdj}  as well as the functional renormalization group method \cite{Pawlowski:2005xe,Dupuis:2020fhh} -- which predict the existence of a critical end point (CEP) that separates the QCD diagram into regions of cross-overs and first-order transitions. 

As the Universe expands and cools down, it follows a certain path in the QCD phase diagram, the so-called \textit{cosmic trajectory}. Assuming a lepton asymmetry of the same order of magnitude as the baryon asymmetry of our Universe ($|Y_L| \sim Y_B$), the transition from the quark-gluon plasma to hadrons should happen at small chemical potentials which implies the cosmic QCD transition to be a cross-over. It has however been shown in a series of papers \cite{Zarembo:2000wj,Schwarz:2009ii,Wygas:2018otj,Middeldorf-Wygas:2020glx,Gao:2021nwz} that allowing for the possibility of a large lepton asymmetry, $|Y_L| \gg Y_B$, shifts the cosmic trajectory towards larger chemical potentials. In particular, the study of \cite{Gao:2021nwz} revealed that \textit{large unequal lepton flavor asymmetries} ($Y_{L_\alpha} \neq \frac{Y_L}{3}$) can render the cosmic QCD transition to be a first-order transition. 
 
\subsection{Cosmic trajectory and state of the art}
\label{sec:Cosmic trajectory and state of the art}

In the following we briefly explain how the cosmic trajectory is calculated in general and summarize the two different methods of \cite{Wygas:2018otj,Middeldorf-Wygas:2020glx} and \cite{Gao:2021nwz}  for the equation of state of QCD that we will compare to.  

 After the electroweak symmetry breaking (at around $T\sim 100$ GeV) and before the onset of neutrino oscillations (at around $T\sim 10$ MeV), the lepton number of individual flavors $L_{\alpha}$, the baryon number $B$ and electric charge $Q$ can be considered as conserved quantities (frozen-out) in the absence of any $B$- or $L$- violating processes. 
 A good approximation to study the evolution of the different charges is to assume that the Universe is an isolated system without any matter or energy exchange.
 Therefore, the total entropy of the thermal bath and particle numbers during the evolution are conserved. Now for the convenience of the computation, the entropy and number density are preferred, and one may define the ratio of these two quantities which then cancels the comoving volume from both and converts the ratio of the original conserved quantities to that of their density counterparts. In other words, $Y_{L_{\alpha}}$, $Y_B$ and $Y_Q$ are conserved quantities. Note that this holds between the phase of entropy dilution (described in sec.~\ref{sec:Late time entropy dilution:} and ~\ref{sec:Entropy dilution}) and the onset of neutrino oscillations and irrespectively of the nature of the transition (i.e. either for a cross-over as well as for a first-order transition).  The cosmic trajectory can be calculated from these five local conservation laws.   

Weak interactions are effective until $T\sim 1$ MeV and justify imposing equilibrium conditions at the timeframe of interest. In that case, each conserved charge is associated with a charge chemical potential, i.e. $\mu_{L_{\alpha}},\mu_B$ and $\mu_Q$, and the cosmic trajectory is simply the solution for $(\mu_{L_{\alpha}},\mu_B, \mu_Q)$ at different temperatures $T$ that keeps $Y_{L_{\alpha}}$, $Y_{B}$ and $Y_Q$ constant\footnote{Out of equilibrium, one would instead have to perform a hydrodynamical simulation to find the solutions for $n_B, n_{L_{\alpha}}$~and $n_Q$. This is often done in the context of heavy ion collisions.}. Let us write out these five conservation laws explicitly, 
\begin{eqnarray}
   Y_{L_\alpha} = & \frac{n_{L_\alpha}}{s}= \frac{n_\alpha + n_{\nu_\alpha}}s \, ,  \quad & \mbox{for\ }\alpha \in \{e,\mu,\tau\}\, , \label{eq:L_alpha}\\
   Y_B = & \frac{n_B}{s}=\sum_{i}\frac{b_i n_i}s \, , \quad & \mbox{with\ }b_i = \rm{baryon\ number\ of\ species\ }i \,, \label{eq:B}\\
   Y_Q = & 0= \frac{n_q}{s}= \sum_{i}{q_i n_i} \, ,          \quad & \mbox{with\ }q_i = \rm{electric\ charge\ of\ species\ }i\, , \label{eq:Q}
\end{eqnarray}
where $n_i$ denotes the net number density (particle minus anti-particle) of particle species $i$, and $s$ is the total entropy density of the Universe. Each particle species on the RHS of Eqs.~\eqref{eq:L_alpha}-\eqref{eq:Q} can be described by a chemical potential $\mu_i$ and a temperature $T_i$. 
Thermal equilibrium implies $T_i=T$ whereas chemical equilibrium implies relations between the chemical potentials of various particle species, e.g. $\mu_u=\mu_c$ (where $u$ and $c$ stand for up and charm quark). Furthermore, the charge chemical potentials  can, in turn, be related to the particle chemical potentials (see e.g. \cite{Schwarz:2009ii} or \cite{Kapusta:2006pm}), i.e.
\begin{equation}
\begin{aligned}
    \mu_{L_{\alpha}} &= \mu_{\nu_{\alpha}}, \\
    \mu_Q &= \mu_u - \mu_d, \\
    \mu_B &= \mu_u + 2 \mu_d.
    \label{eq:charge_chempot}
\end{aligned}
\end{equation}

From now on, we fix the baryon asymmetry to its observed value, $Y_B=8.70\times 10^{-11}$ \cite{Aghanim:2018eyx}, and assume an electric charge neutral Universe by setting $Y_Q=0$. This leaves us with three free input parameters, namely the lepton flavor asymmetries $Y_{L_\alpha}$. The cosmic trajectory is nothing but the solution for the charge chemical potentials $(\mu_{L_{\alpha}},\mu_Q,\mu_B)$ which fulfill Eqs. \eqref{eq:B}-\eqref{eq:Q} at temperatures $T$ and given a choice of $Y_{L_{\alpha}}$.

To calculate the cosmic trajectory, we still have to impose expressions for the net number densities of the particle species and the total entropy density $s$ on the RHS of Eqs. \eqref{eq:B}-\eqref{eq:Q}. The net number density both of leptons and photons are simply momentum integrals over the corresponding thermal distributions (i.e. Fermi-Dirac or Bose-Einstein distributions respectively).
Modeling the QCD sector at temperatures around the QCD transition is however much more complicated since the particle picture is not appropriate due to long-range correlations. The treatment of the QCD sector is also exactly where the two methods of \cite{Wygas:2018otj,Middeldorf-Wygas:2020glx} and \cite{Gao:2021nwz} differ. Let us briefly summarize and compare the lattice QCD based method introduced in \cite{Wygas:2018otj,Middeldorf-Wygas:2020glx} and the functional QCD based method introduced in \cite{Gao:2021nwz}. 

\begin{figure}
    \centering
    \includegraphics[width= \textwidth]{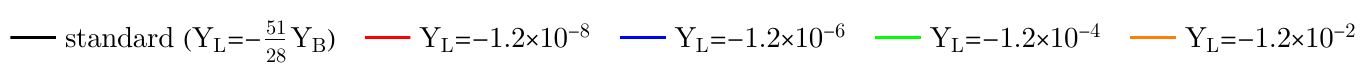}
    \includegraphics[width=0.85 \textwidth]{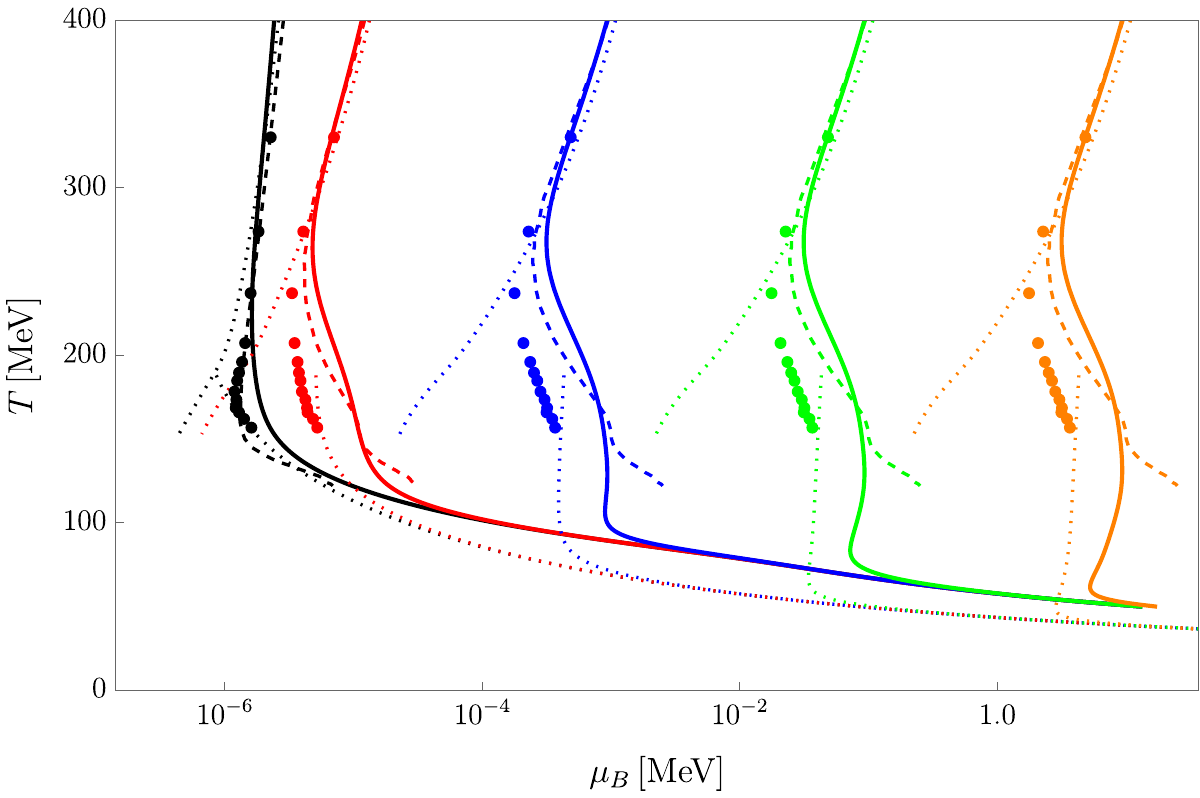}
    \caption{Cosmic trajectories projected onto the $(\mu_B,T)$-plane for different values of the lepton asymmetry $Y_L$ (with $Y_{L_\alpha} = \frac{Y_{L}}{3}$), for the three different methods described in sec.~\ref{sec:Cosmic trajectory and state of the art}: (i) \textit{dotted lines and $\bullet$:} the lattice QCD based method of \cite{Middeldorf-Wygas:2020glx,Wygas:2018otj}, at high $T$ assuming a free quark gas including corrections form perturbative QCD, at intermediate $T$ applying lattice QCD susceptibilities, at low $T$ assuming a HRG approximation; (ii) \textit{dashed lines:} the functional QCD based method of \cite{Gao:2021nwz}, applying the RL truncation; (iii) \textit{solid lines:} the improved functional QCD based method, described in sec. \ref{sec:Update_funcQCD}.}
    \label{fig:funcQCD_comparison}
\end{figure}

The idea behind the method of \cite{Wygas:2018otj,Middeldorf-Wygas:2020glx} is to apply different descriptions of QCD matter in three separate temperature regimes: At high temperatures, quarks and gluons are described as free particles, with \cite{Middeldorf-Wygas:2020glx} including corrections from perturbative QCD \cite{Laine:2006cp,Laine:2015kra}. At low temperatures, a hadron-resonance-gas (HRG) approximation is made, i.e. equilibrium distributions for hadrons are assumed. In the intermediate temperature regime, results from lattice QCD \cite{Bazavov:2014yba,Mukherjee:2015mxc} are applied: To some extent, the sign problem can be circumvented by performing a Taylor expansion of the QCD pressure around zero chemical potentials. At the second order, this introduces so-called susceptibilities which can be calculated on the lattice and through which all thermodynamic quantities (e.g. net number densities) can be expressed. The dotted lines in Fig.~\ref{fig:funcQCD_comparison} show the cosmic trajectory projected onto the $(\mu_B,T)$-plane for different choices of the total lepton asymmetry $Y_L$. Note that we here assumed equal lepton flavor asymmetries ($Y_{L_{\alpha}}= \frac{Y_L}{3}$) simply to facilitate a direct comparison with the works of \cite{Wygas:2018otj} and \cite{Gao:2021nwz}.
As can be seen in Fig.~\ref{fig:funcQCD_comparison}, the use of lattice susceptibilities smoothly connects the high $T$-phase of free quarks with the HRG at low $T$. As pointed out in \cite{Middeldorf-Wygas:2020glx}, the downside of this method is however that the reliability of the Taylor expansion restricts it to relatively small values of the lepton asymmetries. In particular, as a consequence of the sign problem of lattice QCD this method does not allow to explore the possibility of a first-order cosmic QCD transition. 

An alternative method was provided by \cite{Gao:2021nwz}. Instead of relying on lattice QCD, results from functional QCD methods are applied. In particular, at the time of publication of \cite{Gao:2021nwz}, solving the DSEs in the rainbow-ladder (RL) truncation \cite{Gao:2015kea,Isserstedt:2019pgx,Fischer:2018sdj} was the \textit{only QCD based method} delivering not only a complete computation of the QCD phase diagram including the location of a CEP but also the thermodynamic quantities of QCD matter. Including those thermodynamic quantities in the calculation of the cosmic trajectory revealed for the first time that indeed sufficiently large lepton flavor asymmetries can induce a first-order cosmic QCD transition. The dashed lines in Fig.~\ref{fig:funcQCD_comparison} show the cosmic trajectories derived from this second method. The direct comparison between both methods shows however that especially at low temperatures the functional QCD method of~\cite{Gao:2021nwz} does not reproduce the HRG limit.    

\subsection{New functional QCD based method}
\label{sec:Update_funcQCD}

Very recently, the thermodynamic quantities have been calculated from the DSEs with an improved and easily accessed truncation scheme~\cite{Lu:2023msn}. Details about the advances of~\cite{Lu:2023msn} over~\cite{Gao:2015kea} are discussed in the next section sec.~\ref{sec:Improved EoS of QCD}. We included the new results of \cite{Lu:2023msn} into our calculation of the cosmic trajectory and -- as we discuss in section~\ref{sec:Impact on cosmic trajectories} -- find a significantly improved agreement between the fQCD based method and the HRG limit.

\subsubsection{Improved equation of state of QCD}
\label{sec:Improved EoS of QCD}

In general, the QCD equation of state (EoS) at finite temperature $T$ and quark chemical potentials $\boldsymbol{\mu} = \{\mu_q\}$ can be obtained from the logarithm of the partition function using standard thermodynamic relations,
\begin{align}
\frac{P}{T^4} &= \frac{1}{VT^3} \ln Z(T,V,\bm{\mu})\; , \\
n_q &=   \frac{\partial
 P}{\partial \mu_q} \; , \\
 s &=   \frac{\partial
 P}{\partial T} \; , \\
 \epsilon &=  - P+Ts+\mu_q n_q
\; , 
\label{e3pmu}
\end{align}
where $\epsilon$ denotes the total energy density of the Universe. The key elements to compute the number density are the dynamically generated mass of the quark and gluon background condensate $A^a_0$, which are the order parameters of the chiral and deconfinement phase transitions. In principle, they can be directly computed via the gap equation in functional methods under a certain truncation scheme for the DSEs. This has been the foundation of~\cite{Gao:2021nwz}, where the thermodynamic quantities have been calculated for a wide range of chemical potentials and temperatures with the RL truncation~\cite{Gao:2015kea}. The clear disadvantage of this method is unfortunately that the phase transition line derived with the RL truncation does not agree with the lattice results at low chemical potential. Remarkably, the improved truncation scheme of \cite{Gao:2020fbl,Gunkel:2021oya,Fu:2019hdw} leads to a cross-over temperature and phase transition line consistent with the findings of lattice QCD \cite{Borsanyi:2020fev,HotQCD:2018pds,Cea:2014xva}. With this improved truncation the computation of the thermodynamic quantities over a range of chemical potentials and temperatures comparable to \cite{Gao:2021nwz} however becomes computationally very expensive. Therefore, in this work we follow the approach of \cite{Lu:2023msn} and apply an analytic Ising parameterization for the order parameters which incorporates the up-to-date phase transition line from functional methods \cite{Gao:2020fbl,Fu:2019hdw}. The detailed parametrization is given in Ref.~\cite{Lu:2023msn}. 

Following~\cite{Lu:2023msn}, the parameterization applies the phase transition line as:
\begin{equation}
  T_c(\mu_B) = T_c(0) \left[1 - \kappa_{B} \left( \frac{3\mu_{u,d}}{T_c(0)} \right)^2 \right]  \label{eq:TcmuB},
\end{equation}
with $\kappa_{B}=0.016$ , $T_c(0)=155$ MeV. The CEP is found to be located at
\begin{equation}
    (T_{\text{CEP}},\mu_{\text{CEP}})_{u,d} =(118,200) \, \text{MeV} \, .
    \label{eq:CEP}
\end{equation}
Note that the CEP in Eq.~\eqref{eq:CEP} corresponds to a significantly larger chemical potential than the previously applied RL truncation~\cite{Gao:2021nwz,Gao:2015kea}.

On top of this, the order parameter for the deconfinement is also included in the new EoS.
After computing the number density $n(T,\mu)$ for the u/d, s, and c quarks, the entropy density can be expressed as the integral along the chemical potential as
\begin{equation}
\label{eq:pressure}
\delta s(T,\mu)=s(T,\mu)-s(T,\mu=0)=\int_0^\mu d\mu' \, \frac{\partial n(T,\mu')}{\partial T} .
\end{equation}
The full entropy at finite chemical potential  is  given after incorporating the lattice QCD computation at  zero chemical potential with $N_f=2+1+1$ (i.e. $udsc$) as parametrized  in Ref.~\cite{Philipsen:2012nu}, and can be expressed as
\begin{equation}
s_{\rm QCD}=s_{\rm latt}(T,\mu=0)+\delta s(T,\mu).
\end{equation}

Though the analytical form of the here applied Ising parameterization involves an additional zero momentum approximation for the quark propagator (compared to a direct computation from functional methods), this new method is extremely convenient for theoretical studies of heavy-ion-collisions~\cite{Lu:2023msn} or cosmology as in this work. In~\cite{Lu:2023msn}, it was shown that this new method applied to hydrodynamics successfully describes experimental observables like collective flow, particle yields, and particle ratios. 

Following the approach outlined above, the number and entropy densities for the $u,d,s$ and $c$ quark were calculated and tabulated on a grid of $T$ and $\mu_i$ ($i=u,d,s,c$) values, thereby covering the range $50 \leq T \leq 500$ MeV (in $1$ MeV steps) and $0 \leq \mu \leq 1000$ MeV (in $1$ MeV steps). 

\subsubsection{Impact on cosmic trajectories}
\label{sec:Impact on cosmic trajectories}

We extended the C code of \cite{Gao:2021nwz} (itself being a modified version of \cite{Schwarz:2009ii,Wygas:2018otj,Middeldorf-Wygas:2020glx}) in order to include the tabulated thermodynamic quantities described in the previous subsection. The resulting cosmic trajectories are presented in Fig.~\ref{fig:funcQCD_comparison} (solid lines), along with the trajectories derived from the two former methods described in sec.~\ref{sec:Cosmic trajectory and state of the art}. As apparent from Fig.~\ref{fig:funcQCD_comparison}, the new method allows extending the calculation of the cosmic trajectory to smaller temperatures than the old functional QCD based method. Furthermore, the new method reproduces the general behavior of the trajectories as expected at low temperatures, namely $\mu_B$ becoming independent of $Y_L$ and approaching the value of a nucleon mass (see \cite{Schwarz:2009ii}). We conclude that the agreement with the HRG approximation has \textit{significantly} improved. 

Since the new (here applied) method~\cite{Lu:2023msn} predicts the location of the CEP to be at larger values of the u/d chemical potential than the previously applied RL truncation~\cite{Gao:2015kea} --namely $(\mu_{u/d},T)=(200, 118)$ MeV against previously $(\mu_{u/d},T)=(111, 125)$ MeV-- on general grounds we also expect this new method to require larger lepton flavor asymmetries for a first-order cosmic QCD transition than the one reported in \cite{Gao:2021nwz}. In other words, the values in tab. 1 of \cite{Gao:2021nwz} are somewhat underestimated. In the next section, we update the required lepton flavor asymmetries for some of the benchmark scenarios of Tab. 1 in \cite{Gao:2021nwz}. 

Let make a concluding remark on the advantages and disadvantages of the functional QCD based method in comparison to the lattice QCD based method. While with the improved truncation scheme, the functional QCD based method now shows the same generic features in the cosmic trajectory as the lattice QCD based method \cite{Wygas:2018otj,Middeldorf-Wygas:2020glx}, there are still differences between both methods in Fig.~\ref{fig:funcQCD_comparison}. This immediately raises the question of which method is more reliable. For relatively low values of the lepton asymmetries (i.e. for low chemical potentials), where lattice QCD is expected to be exact (within error bars), we certainly advise the use of the QCD based method. The clear disadvantage of this method is however its restriction to low chemical potentials and in particular its limitations to describe a first-order QCD transition at all. While the very existence of a CEP in the QCD phase diagram will only be certain once experimentally confirmed, functional QCD methods generically predict the existence of a CEP and we believe that they are the most reliable tool to consistently describe QCD matter over a large range of temperature and chemical potential. Therefore, when being particularly interested in the phenomenon of a first-order QCD transition, we believe that the functional QCD based method is the best option we have at the moment. 

\subsection{Scan of lepton flavor asymmetries inducing a first-order cosmic QCD transition}
\label{sec:Scan of lepton flavour asymmetries inducing a first-order cosmic QCD transition}

As discussed in sec.~\ref{sec:Symmetry non-restoration and sphaleron freeze-in}, we focus on scenarios with possibly large $Y_{L_{\mu}}$ and $Y_{L_{\tau}}$ but vanishing $Y_{L_e}$. Applying the improved functional QCD method described in sec.~\ref{sec:Update_funcQCD}, we now want to address the question of which values of $Y_{L_{\mu}}$ and $Y_{L_{\tau}}$ we can expect a first order cosmic QCD transition. On general grounds, we expect that a first-order transition happens when the chemical potential of the $u$ and/or $d$ quark become larger than $\mu_{\text{CEP}}$ at $T_{\text{CEP}}$, i.e. when
\begin{equation}
    |\mu_{u/d}| \geq 200 \,  \text{MeV} \hspace{1cm} (\text{at} \, T = 118 \, \text{MeV}) \, .
    \label{eq:first_order_criterion}
\end{equation}
Note that in principle the $c$ and $s$ quark can also experience a first-order phase transition but as their CEP is located at the larger chemical potential this only happens for even larger lepton flavor asymmetries $Y_{L_{\alpha}}$. 

\begin{figure}
    \centering
    \includegraphics[width=0.85 \textwidth]{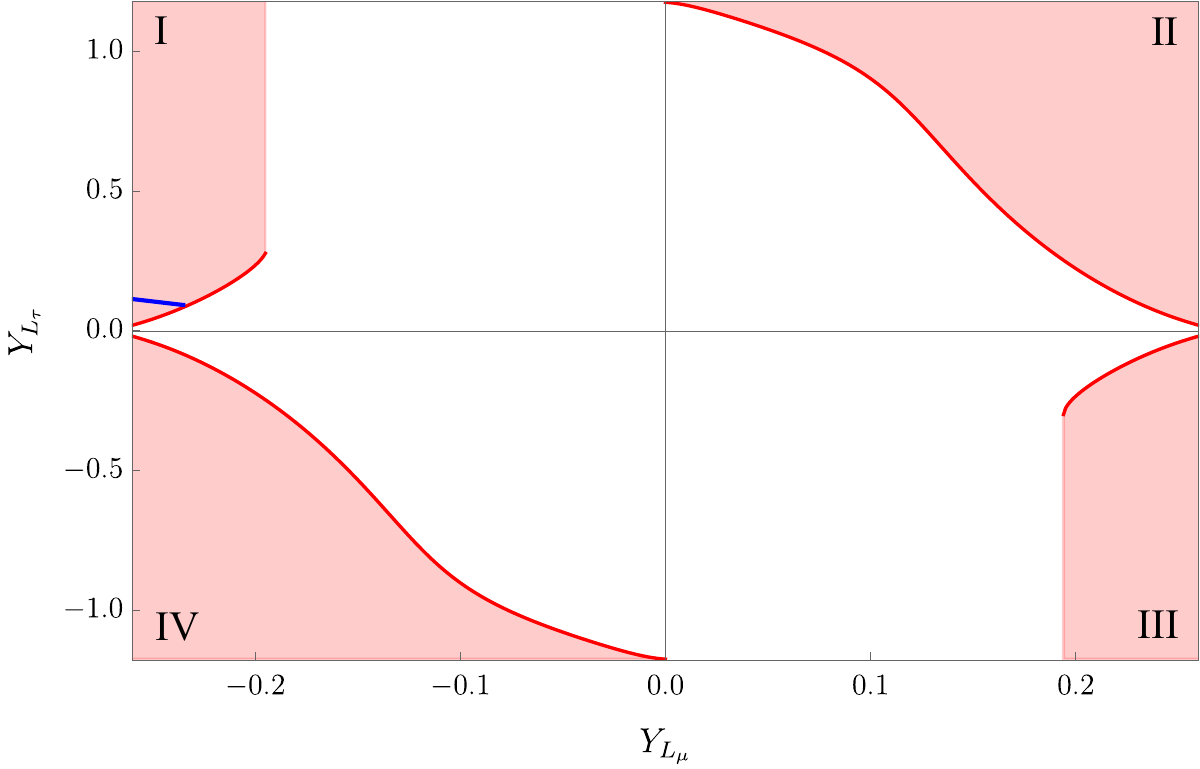}
    \caption{Combinations of  $Y_{L_{\tau}}$- and $Y_{L_{\mu}}$-values which induce a first-order cosmic QCD transition (shaded regions). The blue line marks the benchmark region (Eq.~\eqref{eq:benchmark_line}) which also leads to successful baryogenesis assuming an entropy dilution of $\Delta=18$. Note that no numerical solutions were found for the borders without solid red lines.}
    \label{fig:CEPScan}
\end{figure}

We implemented Eq. \eqref{eq:first_order_criterion} (at the equality) as an additional equation on top of the conservation laws in Eqs.~\eqref{eq:L_alpha}-\eqref{eq:Q} into our C code. This reduces the number of degrees of freedom of the system of equations by one. To be more specific, in general only two of the $Y_{L_{\alpha}}$ are now free input parameters while the third one is obtained as a solution to Eqs.~\eqref{eq:L_alpha}-\eqref{eq:Q} and \eqref{eq:first_order_criterion}. Keeping $Y_{L_e}=0$ fixed, we therefore vary $Y_{L_{\mu}}$ and numerically solve Eqs.~\eqref{eq:L_alpha}-\eqref{eq:Q} for $T=118$ MeV together with Eq.~\eqref{eq:first_order_criterion}, obtaining thereby solutions for $Y_{L_{\tau}}$. Using the solution as an adaptive guess for the next $Y_{L_{\mu}}$ value speeds up the computation time of the scan significantly. It turns out that $|\mu_u|$ is always slightly larger than $|\mu_d|$, such that Eq.~\eqref{eq:first_order_criterion} for the $u$ quark is a sufficient criterion for a first-order transition. The result of our scan for $(Y_{L_\mu}, Y_{L_\tau})$-values is presented as the red lines in Fig.~\ref{fig:CEPScan}. We will discuss in the next section \ref{sec:Lensing effect of QCD critical end point} why this line should only be understood as an estimate. In particular, we found that in some cases already slightly smaller values induce a first-order transition. 
The general behavior of the line in the regions of either $(Y_{L_{\mu}},Y_{L_\tau})>0$ (II) or $(Y_{L_{\mu}},Y_{L_\tau})<0$ (IV) is relatively easy to understand: For a larger $Y_{L_{\mu}}$ a smaller $Y_{L_{\tau}}$ is already sufficient to induce a first-order transition, and vice versa. The figure also shows that in general $Y_{L_{\mu}}$ is more efficient in inducing a first-order transition than $Y_{L_{\tau}}$: For $Y_{L_{\mu}}=0$ a value of $|Y_L|=|Y_{L_{\tau}}| \gtrsim 1.2$ is required while for $Y_{L_{\tau}}=0$ a value of $|Y_L|=|Y_{L_{\mu}}| \gtrsim 0.3$ is already sufficient. For the regions $(Y_{L_{\mu}}>0,Y_{L_\tau}<0)$ (III) and $(Y_{L_{\mu}}<0,Y_{L_\tau}>0)$ (I), it even turns out that a larger $Y_{L_{\tau}}$ cannot compensate a small $Y_{L_{\mu}}$ any longer if $|Y_{L_{\mu}}| \lesssim 0.2$. Note that in general the cosmic trajectories in the regions I and III are less monotonic than the trajectories in the regions II and IV. This is because not only the individual lepton flavor asymmetries $Y_{L_{\alpha}}$ matter but also the total lepton asymmetry $Y_L$. In particular, it has been shown in \cite{Middeldorf-Wygas:2020glx} that at high temperatures (when the ultra-relativistic limit applies) the quark chemical potentials become independent of the individual $Y_{L_{\alpha}}$ and only depend on the total asymmetry $Y_L$~\footnote{Hence, the case of $Y_{L_{\mu}}=-Y_{L_{\tau}}$ e.g. is only able to induce a first-order transition due to the non-trivial impact of the quark masses.}.
Since in regions II and IV enhancing $|Y_{L_{\mu}}|$ at fixed $Y_{L_{\tau}}$ (and vice versa) always implies enhancing $Y_L$, it also directly implies enhancing the quark chemical potentials. The same does however not hold in the regions I and III, where one also has to regard the value of the total lepton asymmetry $Y_L$. 
It is therefore not surprising that the shaded region in Fig.~\ref{fig:CEPScan} for regions I and III is not symmetric to regions II and IV.   

A comparison of our scan with the results of sec.~\ref{sec:Symmetry non-restoration and sphaleron freeze-in} shows that certain combinations of $Y_{L_{\mu}}$ and $Y_{L_{\tau}}$ values not only lead to the observed value of the baryon asymmetry but also induce a first-order cosmic QCD transition. 
We highlight a combination of $(Y_{L_\mu},Y_{L_\tau})$ by the blue line in Fig.~\ref{fig:CEPScan} and refer to them as our benchmark scenarios for the rest of this work. The analytical expression for this benchmark line is
\begin{equation} 
Y_{L_\tau} = -0.11 - 0.86 \cdot Y_{L_\mu}  
    \label{eq:benchmark_line}
\end{equation}
and we expect first-order transitions for $Y_{L_\mu} \lesssim - 0.23$. 

Last but not least, let us compare the findings of the scan described in this section with some of the values reported in Tab.~1 of \cite{Gao:2021nwz} based on the RL truncation. 
In particular, scenario (ii) with $(Y_{L_e=0},Y_{L_\mu}=-Y_{L_\tau})$ was reported to require $Y_{L_\mu}>7.4 \times 10^{-2}$ whereas the scan in Fig.~\ref{fig:pot_genesis} finds $Y_{L_\mu}>2.0\times 10^{-1}$. Another scenario of interest (though not represented by Fig.~\ref{fig:pot_genesis}) may be one of equal lepton flavor asymmetries, i.e. $Y_{L_e}=Y_{L_\mu}=Y_{L_\tau}=\frac{Y_L}3$: The method based on the RL truncation found $Y_L> 1.1\times 10^{-1}$~\cite{Gao:2015kea} in contrast to $Y_L> 3.5 \times 10^{-1}$ from the improved method applied in this work. As discussed in sec.~\ref{sec:Update_funcQCD}, this underestimation (in both cases by a factor of $\sim 3$) by \cite{Gao:2021nwz} is due to the smaller chemical potential of the CEP when applying the RL truncation.

\subsection{Lensing effect of QCD critical end point}
\label{sec:Lensing effect of QCD critical end point}

\begin{figure}
    \centering
    \includegraphics[width=0.85\textwidth]{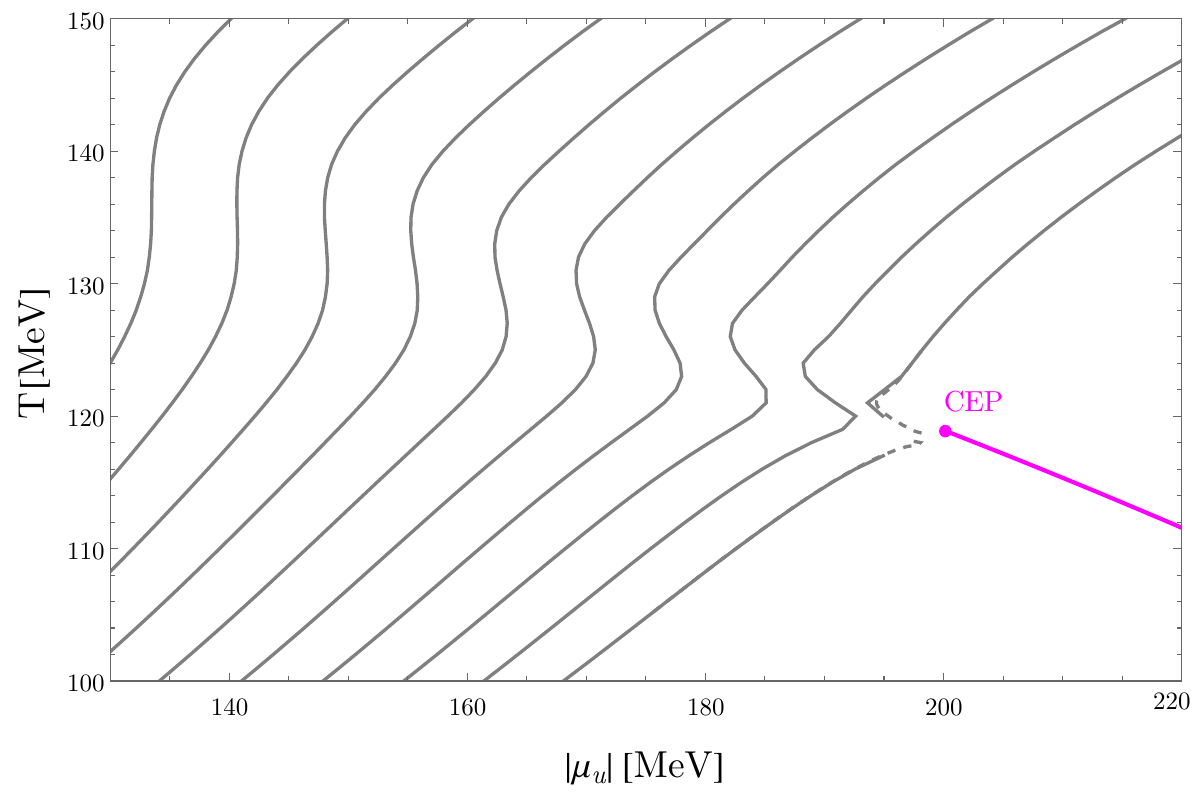}
    \caption{Lensing effect of QCD: Cosmic trajectories with different values of $Y_{L_{\mu}}$ and $Y_{L_{\tau}}$ (chosen along our benchmark line in Eq. \eqref{eq:benchmark_line}) get deformed towards the CEP. The dashed curve has been obtained with a finer temperature grid in the numerical tables described in sec.~\ref{sec:Improved EoS of QCD}.}
    \label{fig:focussing_effect}
\end{figure}

In this section, we elaborate on an interesting feature of the cosmic trajectories, namely the bending of trajectories towards the CEP when passing through its vicinity. This phenomenon is known as the \textit{critical lensing or focussing effect of QCD} and has been reported in the context of heavy-ion collisions \cite{Asakawa:2008ti,Dore:2022qyz,Nonaka:2004pg}. In Fig.~\ref{fig:focussing_effect} we show several cross-over trajectories in the ($\mu_u,T$)-plane with different values of $Y_{L_\mu}$ and $Y_{L_\tau}$. The values of $Y_{L_\mu}$ and $Y_{L_\tau}$ are chosen such that they are an extension of our benchmark line in Eq.~\eqref{eq:benchmark_line} into the cross-over region. Since the CEP behaves as an attractor, we indeed observe how the trajectories get deformed into the direction of the CEP for increasing values of $Y_{L_\mu}$. 

The same behavior is also the reason why our code is currently not capable of calculating first-order transition trajectories in the CEP region. As explained e.g. in \cite{Asakawa:2008ti,Dore:2022qyz,Nonaka:2004pg}, since the trajectories merge towards the CEP, the density of trajectories with different values of $Y_{L_{\mu}}$ (and $Y_{L_{\tau}}$) passing through the same region is expected to be very high near the CEP.
Resolution of the different trajectories in the CEP region would therefore require extremely high numerical precision and in particular a very fine spacing of $\mu_{u,d}$ in the numerical tables described in sec.~\ref{sec:Improved EoS of QCD}. The authors of \cite{Dore:2022qyz} e.g. assumed a grid size of $\Delta \mu_{u,d}=0.1$ MeV\footnote{private communication.}, in contrast to $\Delta \mu_{u,d} = 1$ MeV as in our work. For the same reason, the first-order transition line we found from the scan in sec. \ref{sec:Scan of lepton flavour asymmetries inducing a first-order cosmic QCD transition} should rather be understood as an estimate. For a restricted temperature range of $110<T<125\,$~MeV we have decreased the temperature spacing in the numerical tables described in sec.~\ref{sec:Improved EoS of QCD} to $\Delta T=0.1$ MeV, which helped to get somewhat closer to the CEP region; in fact, the trajectory with $Y_{L_{\mu}}=-0.22$ in Fig.~\ref{fig:focussing_effect} has been calculated with this reduced $\Delta T$ (dashed lines) which helped to somewhat extend the temperature range where solutions can be found. However, as we show in the next section, for first-order transition trajectories the real limitation is not the temperature spacing $\Delta T$ (which does not really enhance the numerical precision of our calculation but only helps with the adaptive guesses for the solutions) but the spacing in the chemical potential $\Delta \mu_{u,d}$. While reducing $\Delta \mu_{u,d}$ in our numerical tables (sec. \ref{sec:Improved EoS of QCD}) is in principle possible, it goes beyond the scope of this work due to the related raise in computation time to produce the numerical tables described in sec. \ref{sec:Improved EoS of QCD}. However, in the next section, we demonstrate how we can estimate the GW spectrum expected from a first-order transition and how this can be related to values of $Y_{L_{\mu}}$ and $Y_{L_{\tau}}$.


\section{Gravitational waves}
\label{sec:Gravitational waves}

A cosmological first-order phase transition can provide a stochastic gravitational wave background that can be observable today \cite{Winicour:1973ApJ, Athron:2023xlk,Caprini:2010xv,NANOGrav:2021flc,Xue:2021gyq,DiBari:2021dri,Madge:2023dxc,Roshan:2024qnv,Guo:2024gmu}. The possibility of a first-order QCD transition is a key motivation behind the $\mu$-Ares proposal \cite{Sesana:2019vho}, which will deploy three interferometers at the Lagrange points of the orbit of the Earth around the sun.  In Ref.~\cite{Gao:2023djs}, we pointed out how such a signal can potentially arise from a strong cosmological first-order QCD phase transition.
There are few scenarios in which the QCD transition can be made first order, though as mentioned previously, the possible catalyst of a large lepton asymmetry has been known for over a decade \cite{Schwarz:2009ii,Wygas:2018otj}. 
The gravitational wave spectrum from a first-order transition has three contributions. A collision term due to the colliding scalar shells, an acoustic term due to the sound waves, and a turbulence term. It is generally accepted that the acoustic term dominates \cite{Caprini:2019egz}, which we will correspondling assume in the following. 

\subsection{Basic expressions of the gravitational wave spectrum}
  An accurate gravitational wave spectrum from a first-order transition requires a simulation. A reasonable approximation can be made from a semi-analytic model known as the sound shell model \cite{Hindmarsh:2015qta,Hindmarsh:2017gnf} -- where the gravitational wave spectrum is approximated by an incoherent superposition of the contribution of many individual sound shells. Here, the sound shell refers to the plasma shell that is the part of the plasma driven from equilibrium wherein the majority of the energy released in the phase transition is dumped into. The result of the sound shells is a gravitational power spectrum, whose peak wave number corresponds to the mean bubble separation. As gravitons are very weakly interacting, the spectrum has no further evolution after its formation apart from redshifting like radiation. We can therefore measure the abundance, $\Omega _{\rm GW}$ of gravitational radiation as a function of frequency, $f$, and infer that we have a well-preserved photograph of an event in the early Universe. The spectrum we would see today, assuming the sound shell model, has a parameter-independent shape which goes as $f^{-4}$ in the UV, $f^3$ (from causality) in the infrared, and has a brief plateau in between. The position of this plateau/peak in the frequency/amplitude plane are the only two observables,\cite{Hindmarsh:2013xza,Hindmarsh:2015qta,Hindmarsh:2017gnf,Guo:2020grp} \footnote{Simulations suggest a doubly broken power law due to there being a second scale in the problem - the thickness of the sound shell \cite{Gowling:2021gcy}. Since modeling of this effect currently does not exist, we ignore this effect and use the standard analytic fits. }
The spectrum depends upon the bubble wall velocity $v_w$, the duration of the transition  $\beta ^{-1}$ and the average fluid velocity $\bar{U}_f$, which reads \cite{Hindmarsh:2013xza,Hindmarsh:2015qta,Hindmarsh:2017gnf,Guo:2020grp}:
\begin{eqnarray}
h^2 \Omega_{\text{GW}} =  1.2 \times 10^{-6} \left(\frac{100}{g_{\ast}(T_\ast)}\right)^{1/3} \Gamma^2 \bar{U}_f^4  
\left[
\frac{H_s}{\beta(v_w)}
\right]
v_w 
\times
\Upsilon S_{\rm sw}(f),
\end{eqnarray}
where $g_*$ is the relativistic degrees of freedom, $H_s$ is Hubble rate. ${\Upsilon}$ accounts for the fact that the lifetime of  the soundwaves, $\tau_{\text{sw}}$, is finite. For radiation domination, we have
\begin{equation}
\Upsilon = 1 - \frac{1}{\sqrt{1 + 2 \tau_{\text{sw} }H_s}} .
\end{equation}
The adiabatic index is given by $\Gamma \sim 4/3$, and 
the root mean squared (RMS) fluid velocity is determined by the PT strength $\alpha$, $\bar{U}_f^2\sim \frac{3}{4} \kappa _f (\alpha) \alpha $ and here we set as
\begin{equation}
\bar{U}_f^2= \frac{3}{4} \frac{\kappa \alpha}{1 + \alpha},
\end{equation}
with PT strength $\alpha = \theta/\rho _{\rm rad}$ where $\theta \equiv \Delta p-\frac{1}{4}(T \Delta s+\mu_q \Delta n)$ is the trace anomaly  with $(\Delta p,\Delta s, \Delta n)$  the pressure, entropy, and density differences between the phases respectively  and $\rho _{\rm rad}$ is the radiation energy density defined as 
$\rho _{\rm rad}=\pi^2/30 g_* T_*^4$.

The spectral form has the shape
\begin{eqnarray}
S_{\rm sw}(f) = \left( \frac{f}{f_{\rm sw}} \right)^3 \left( \frac{7}{4+3(f/f_{\rm sw})^2} \right)^{7/2}    
\end{eqnarray}
and the peak frequency is controlled by the temperature of the transition, $T_\ast$, the bubble wall velocity and the duration of the transition
\begin{equation}
    f_{\rm sw} = 6.23 \times 10^{-6} {\rm Hz} \frac{1}{v_w} \left( \frac{\beta}{H_s} \right) \left( \frac{T_\ast}{100 {\rm GeV}} \right) \left( \frac{g_\ast}{100}\right)^{1/6} \ .
\end{equation}

\subsection{Gravitational wave spectrum  and detection prospects}
\label{sec:Transition temperatures for benchmark scenarios}
\begin{figure}
    \centering
    \includegraphics[width=0.85\textwidth]{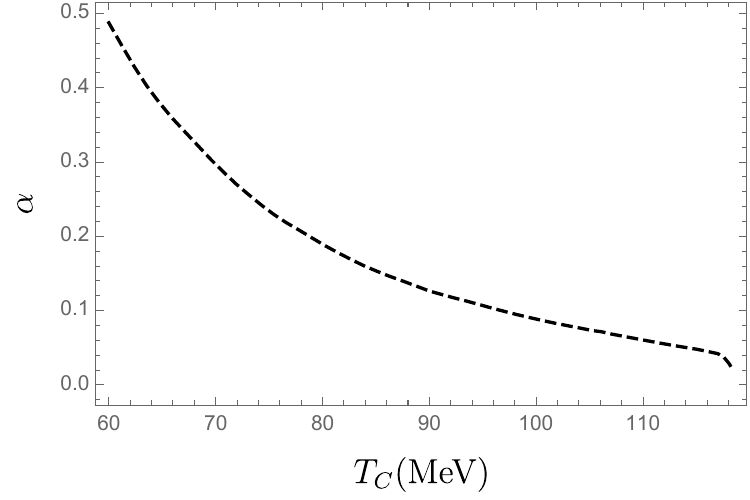}
    \caption{The minimum value of $\alpha$ as a function of the critical temperature, neglecting the contribution of the chemical potential to the energy density of the Universe.}
    \label{fig:tcvalpha}
\end{figure}

 Calculating the thermal parameters $\beta $ and $v_w$ is quite challenging, as it is difficult to get the thermodynamic potential non-perturbatively as in the QCD phase transition. However, the current development based on the functional QCD approaches has made it possible to access the equation of state of QCD in the whole plane of temperature and chemical potential as mentioned above. The trace anomaly can be then extracted from the EoS as:
\begin{equation}
    \theta = \Delta p - T_c 
\frac{1}{4} \Delta s-\mu_q\frac{1}{4} \Delta n  \geq \left.\left(- T_c 
\frac{1}{4} \Delta s-\mu_q\frac{1}{4} \Delta n\right) \right| _{T_c} .
\end{equation}
Here $T_c$ is the critical temperature at which the pressure difference between the phases vanishes, $(\Delta p,\Delta s, \Delta n)$ are the pressure, entropy, and density differences between the phases respectively. We follow the above argument and assume the first-order phase transition to take place at approximately $\Delta p=0$ (implying percolation temperature $T_*\sim T_c$), and the entropy and density differences  $(\Delta s, \Delta n)$ can be directly read off from the EoS at the two sides of the PT line. Using the method discussed in sec.~\ref{sec:Improved EoS of QCD}, we can therefore calculate $\alpha$ as a function of the critical temperature in the first- order phase transition region, which we plot in Fig.~\ref{fig:tcvalpha}. Note that in our calculations of the trajectories, we assume to be near equilibrium. If there is a substantial amount of supercooling, this assumption breaks down. However, we expect the surface tension to be quite small and therefore there to be only a modest amount of supercooling, see Appendix.~\ref{app:Note on equilibrium}. Hence, we can estimate the value of the trace anomaly from our calculation of the trajectories.  For the range of values for $\theta$ that we consider, the signal today has an approximate scaling of $\Omega _{\rm GW}h^2 \sim 10^2 \theta ^3 v_w/\beta ^2$ (for details and exact expressions see \cite{Guo:2020grp}). Therefore, calculating $\theta$ under the assumption of limited supercooling is a conservative estimate.

\begin{figure}[t]
    \centering
    \includegraphics[width=0.85\textwidth]{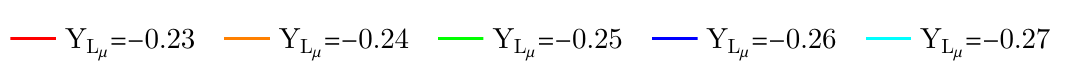}
    \includegraphics[width=0.85 \textwidth]{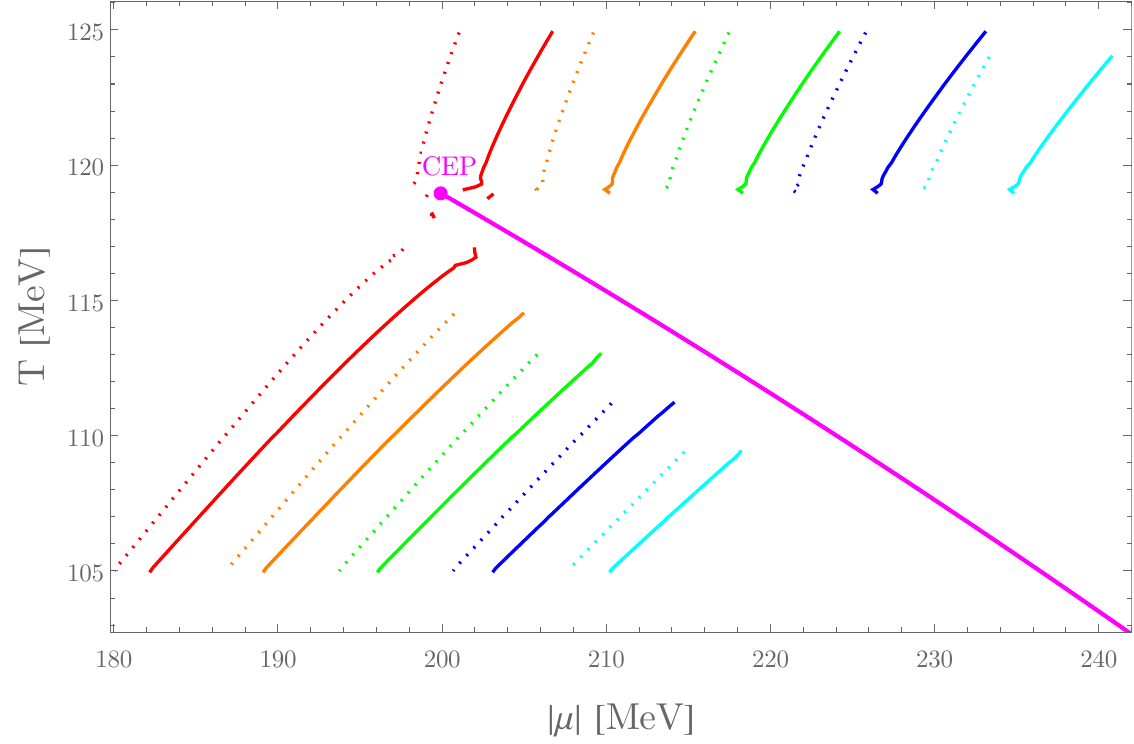}
    \caption{Cosmic trajectories around $T_c$ in the $(\mu_u,T)$-plane (solid) and $(\mu_d,T)$-plane (dotted) for different values of $Y_{L_{\mu}}$ along our benchmark line in Eq. \eqref{eq:benchmark_line}.}
    \label{fig:Ts_benchmarks}
\end{figure}

Under the assumption of no supercooling, the peak frequency of the GW spectrum will be controlled by the critical temperature $T_c$. As we explained in sec.~\ref{sec:Lensing effect of QCD critical end point}, the exact prediction of $T_c$ with our technique is at the moment not feasible due the high level of numerical precision required due to the lensing effect of QCD. Fig.~\ref{fig:Ts_benchmarks} shows several first-order trajectories in the $(\mu_{u/d},T)$-plane along our benchmark line in Eq.~\eqref{eq:benchmark_line}. All trajectories show a gap where our code does not manage to find solutions. From the discussion of sec.~\ref{sec:Lensing effect of QCD critical end point}, we expect that within this temperature range, the trajectories merge towards the CEP resulting in a very high density of trajectories in that region.  
While it is not clear at which temperatures the individual trajectories would cross the phase transition line Eq.~\eqref{eq:TcmuB}, it is immediately clear that this crossing happens within the temperature gap. The benchmark scenarios displayed in Fig.~\ref{fig:Ts_benchmarks} therefore refer to transition temperatures within the range $T_c=109-118$~MeV,
\begin{equation}
    Y_{L_{\mu}} \in [-0.23, -0.27 ] \,  \Rightarrow \, T_c \in [109, 118] \, \text{MeV}  \, .
    \label{eq:benchmark_Ts}
\end{equation}

\begin{figure}[t]
    \centering
    \includegraphics[width=0.85\textwidth]{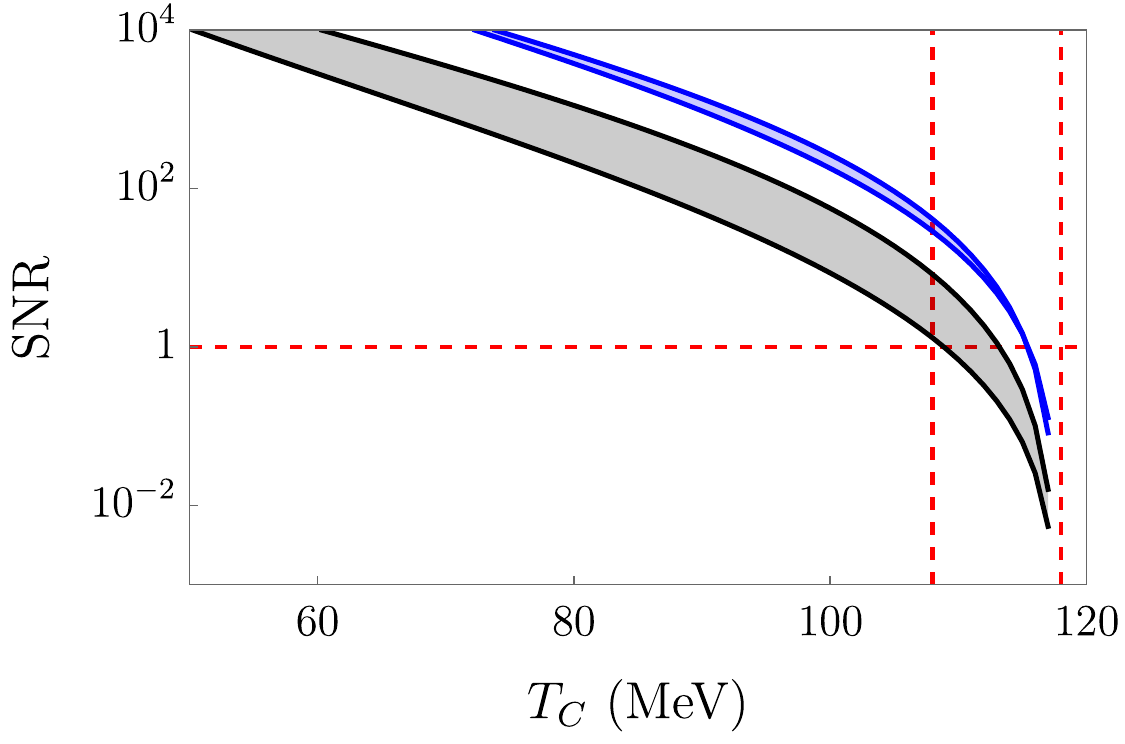}
    \caption{The minimum signal-to-noise ratio of a first-order QCD phase transition as a function of the critical temperature $T_c$. The gray and blue band corresponds to the cases of entropy dilution $\Delta=18$ and $\Delta=1$ (no entropy dilution), respectively. We restrict the phase transition time scale to be between $10^{(2-4)}$, which corresponds to $T_c \in [109, 118] \, \text{MeV}$, indicated by vertical red dashed lines.}
    \label{fig:SNR}
\end{figure}

\begin{figure}
    \centering
\includegraphics[width=\textwidth]{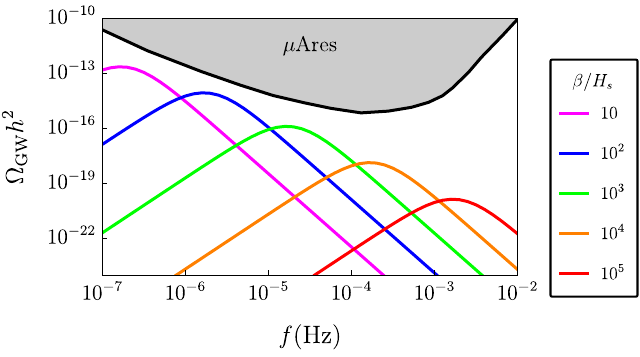}
\includegraphics[width=\textwidth]{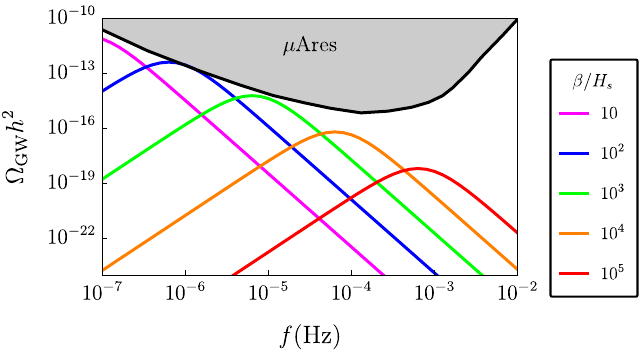}
    \caption{Gravitational wave spectra with thermal parameters $(\alpha, v_w, T_c) = (0.1,1,0.12~{\rm GeV} )$ for different values of  inverse transition timescales. We show the case of an entropy dilution of $\Delta=18$ (top) and the case of no entropy dilution (bottom). The sensitivity of $\mu$Ares is only weakly sensitive to the timescale of the transition ($\beta/H_s$) for a reasonably typical range.}
    \label{fig:GWbenchmark}
\end{figure}

Such a temperature range then yields the parameter $\alpha$ as:
\begin{equation}
   \alpha= [10^{-3},0.078] \, ,
    \label{eq:alpha}
\end{equation}
Of course, in general, larger values of $Y_{L_{\mu}}$ would lead to smaller values of $T_c$. We simply chose the values in Eq.~\eqref{eq:benchmark_Ts} to have some reference values in the remainder of this work. As can be seen in Fig.~\ref{fig:Ts_benchmarks} and as anticipated in sec.~\ref{sec:Scan of lepton flavour asymmetries inducing a first-order cosmic QCD transition}, for all scenarios we find slightly larger values for $|\mu_u|$ than for $|\mu_d|$. The difference between $|\mu_u|$ and $|\mu_d|$ is however so small that -- except for a very small range of $(Y_{L_{\mu}},Y_{L_\tau})$-values -- one can safely assume that if the $u$ quark experiences a first-order transition the $d$ quark experiences a first-order transition, too. This implies an additional factor of $2$ for the trace anomaly. We do note, however, that we are neglecting the chemical potential contribution to the energy density of the Universe when computing the parameter $\alpha$, which would reduce the size of the gravitational wave density as $\alpha \propto 1/\rho _{\rm rad}$. We expect this to be safe at the relevant temperature range, where $\mu_{q,L_\alpha,\nu_\alpha}/(2\pi T)\lesssim 1$, and hence, the chemical potential contributes less than the temperature in the energy density.

Let us turn our attention to the remaining thermal parameters that determine the GW spectrum. The gravitational wave spectrum has a modest dependence on the bubble wall velocity. However, there is an indirect effect through the efficiency factor $\kappa$ which can take on a different dependence on $\alpha$ depending on how the wall velocity compares to the speed of sound. However, recent work seems to suggest that it is difficult to avoid a scenario where the wall velocity is supersonic \cite{Krajewski:2024gma}. This is a regime where the dependence on the wall velocity is very weak, so for simplicity, we take $v_w=1$. There is, unfortunately, no good guess for the timescale of the transition except to notice for first-order phase transitions $10^2 \lesssim \beta/H_s \lesssim 10^4$ seems to be a reasonable phenomenologically motivated range \cite{Caprini:2019egz}. To calculate the signal-to-noise (SNR) ratio of a hypothetical spectrum observed by $\mu$Ares, we use the data from \cite{Sesana:2019vho} and note that
\begin{equation}
    {\rm SNR} = \sqrt{T \int _{f _{\rm min}} ^{f_{\rm max}} \left( \frac{\Omega _{\rm GW}(f) }{N(f)} \right) ^2}
\end{equation}
where $N(f)$ is the noise function of the detector and $T$ is the lifetime of an experiment which we take to be one year. \par

Surprisingly, the signal-to-noise ratio is relatively independent of $\beta/H _s$ as seen in Fig.~\ref{fig:SNR}. This can be understood by noting that the peak sensitivity of $\mu$Ares is slightly higher than the frequency of the peak amplitude of the gravitational wave from a strong QCD transition. Shorter transitions have a smaller peak amplitude, which is compensated by the fact that the position of the peak moves closer to the peak sensitivity as shown in Fig.~\ref{fig:GWbenchmark}. We note that the choice of the wall velocity here offers an upper limit of the estimation.  A recent analysis~\cite{Cline:2025bwe} that applies the same EoS of QCD together with an earlier calculation of the surface tension of QCD~\cite{Gao:2016hks} shows that the QCD first order phase transition is weak and the wall velocity is much smaller than unity.  This could greatly suppress the gravitational wave signals and could make the QCD phase transition out of reach of  the next generation detector of $\mu$Ares.  Furthermore, a self consistent determination of the wall velocity and also the GW parameter $\beta$ is still required  which can be achieved  by directly calculating the QCD effective potential~\cite{Zheng:2023tbv,Zheng:2024tib}. A determination of the surface tension and the description of the spinodal line of QCD is currently in progress.

\section{Conclusions}
\label{sec:Conclusions}
We have explored the scenario where substantially large lepton asymmetries can generate the observed baryon asymmetry of the Universe. By preventing the restoration of electroweak symmetry, these asymmetries effectively suppress the sphaleron rate, leading to a mechanism of baryogenesis via an exponentially suppressed rate of conversion from lepton to baryon asymmetry, termed sphaleron freeze-in. Contrary to previous perturbative method based estimates, our use of dimensional reduction techniques showed that the required lepton asymmetries for successful baryogenesis are an order of magnitude smaller than previously estimated.

A large lepton asymmetry can also induce a first-order cosmic QCD transition, a phenomenon accompanied by the potential emission of GWs. Although challenges remain, particularly regarding the precise location of the QCD critical endpoint, recent advancements in functional QCD and improved truncation schemes provide promising avenues for refining our understanding of these cosmic transitions. We identified a parameter range where $ Y^{\text{ini}}_{L_\mu}=-Y^{\text{ini}}_{L_\tau} $, capable of triggering a first-order cosmic QCD transition, aligns with constraints from the CMB and BBN (constraining the lepton asymmetry at the CMB temperature, predominantly stored in the form of Dirac neutrino flavor asymmetries) without requiring entropy dilution. Nevertheless, achieving both successful baryogenesis and a first-order QCD transition necessitates minimal entropy dilution between the QCD transition and BBN epochs.

Given that the GW signal from a strong first-order transition is generally dominated by a contribution arising from sound waves, we used the sound shell model with the difference in the trace anomaly between the two phases extracted from the EoS of QCD to explore the potential GW signals. This demonstrated the potential for upcoming GW experiments, such as $\mu$Ares, to probe and potentially validate the sphaleron freeze-in paradigm. Decreasing the transition time relative to the Hubble time reduces the maximum gravitational wave (GW) amplitude while simultaneously raising the peak frequency, bringing the spectrum nearer to the optimal sensitivity range of $\mu$Ares.

 Addressing fundamental questions about the magnitude of lepton asymmetry required for successful baryogenesis, the conditions for inducing a first-order cosmic QCD transition, and the GW signals from such events, our study contributes to a comprehensive framework to explore the large lepton asymmetry. Further insight into the surface tension of bubbles within the hadronic phase would be pivotal for further enhancing predictions for GW signals from the QCD transition.

\acknowledgments
FG is supported by the National  Science Foundation of China under Grants  No. 12305134. JH acknowledges support from the Cluster of Excellence “Precision Physics, Fundamental Interactions, and Structure of Matter” (PRISMA$^+$ EXC 2118/1) funded by the Deutsche Forschungsgemeinschaft (DFG, German Research Foundation) within the German Excellence Strategy (Project No. 390831469). CH acknowledges support from the CDEIGENT grant No. CIDEIG/2022/16 (funded by Generalitat Valenciana under Plan Gen-T), and from the Spanish grants PID2020-113775GB-I00 (AEI/10.13039/501100011033) and Prometeo CIPROM/2021/054 (Generalitat Valenciana). JH and CH also acknowledge support from the Emmy Noether grant "Baryogenesis, Dark Matter and Neutrinos: Comprehensive analyses and accurate methods in particle cosmology" (HA 8555/1-1, Project No. 400234416) funded by the Deutsche Forschungsgemeinschaft (DFG, German Research Foundation). YL is supported by the National  Science Foundation of China under Grants  No. 12175007 and No. 12247107. IMO acknowledges support by Fonds de la recherche scientifique (FRS-FNRS).

\newpage
\appendix 
\textbf{\large{Appendix}}
\section{Effective potential at finite chemical potential}\label{app:effpot}
In this appendix, we present the main relations and expressions useful for studying the full 4-dimensional electroweak theory at a finite lepton number density using a 3-dimensional effective theory. The presence of a finite fermion number density, on one hand new corrections arise in the renormalization of the fields and parameters upon integrating out the heavy modes, as compared to the case of vanishing chemical potentials. On the other hand, new terms arise in the dimensionally reduced effective theory due to the reduction of symmetries in the original 4-dimensional theory in the presence of finite chemical potentials.

 We will denote the coefficients of the terms present already for vanishing chemical potential by the subscript $\mu=0$. The finite chemical potential corrected Debye masses are given by
\begin{eqnarray}
m_D^2 & = & m_{D,\mu=0}^2 + \frac{g^2}{4\pi^2}\left(\mu_B^2 + \sum_{i=1}^{n_f} \mu_{L_i}^2\right),  \\
m_D^{\prime 2} & = & m_{D,\mu=0}^{\prime 2} + \frac{{g'}^2}{4\pi^2}\left(\frac{11}{9}\mu_B^2 + 3\sum_{i=1}^{n_f} \mu_{L_i}^2 \right).
\end{eqnarray}
The zero chemical potential Debye masses are given by
\begin{eqnarray}
m_{D,\mu=0}^2 & = & \biggl(\frac{2}{3}+\frac{N_s}{6}+\frac{n_F}{3}\biggr) g^2 T^2 \, ,  \\
m_{D,\mu=0}^{\prime 2} & = & \biggl(\frac{N_s}{6}+\frac{5n_F}{9}\biggr) g'^2 T^2 \, ,
\end{eqnarray}
where $n_F=3$ is the number of families and $N_s=1$ is the number of Higgs doublets.

The finite chemical potential corrected coupling coefficients for the terms already present for $\mu=0$ are given by
\begin{eqnarray}
\lambda_3 & = & \lambda_{3,\mu=0} -\frac{3g_Y^2T}{16\pi^2}\left(g_Y^2 - 2\lambda\right){\cal A}\left(\frac{\mu_B}{3}\right), \\
h_3 & = & h_{3,\mu=0} + \frac{g^4T}{192\pi^2}\left(9{\cal A}\left(\frac{\mu_B}{3}\right) + \sum_{i=1}^{n_f} {\cal A}\left(\mu_{L_i}\right)\right),  \\
g_3^2 & = & g_{3,\mu=0}^2 +\frac{g^4T}{48\pi^2}\left(9{\cal A}\left(\frac{\mu_B}{3}\right) + \sum_{i=1}^{n_f} {\cal A}\left(\mu_{L_i}\right)\right),
\end{eqnarray}
where
\begin{equation}
{\cal A}(\mu) = \psi\left(\frac{1}{2}+\frac{i\mu}{2\pi T}\right) + \psi\left(\frac{1}{2}-\frac{i\mu}{2\pi T}\right) + 2\gamma_E + 2\ln\,4.
\end{equation}
with $\psi(z)\equiv \partial_z \ln \Gamma(z)$. The corresponding vanishing chemical potential ($\mu=0$) coupling coefficients are given by
\begin{eqnarray}
\lambda_{3,\mu=0} & = & \lambda(\mu)T\biggl\{
1-\frac{3}{4}\frac{g^2}{16\pi^2}\biggl[\biggl(
\frac{6}{h^2} -6+2h^2 \biggr)L_b+
\biggl(4 t^2-8\frac{t^4}{h^2}\biggr)L_f-
\frac{4}{h^2} \biggr]\biggr\} \, ,\\
h_{3,\mu=0} & = & \frac{1}{4}g^2(\mu)T
\biggl[1+ \frac{g^2}{16\pi^2}\biggl(
\frac{44-N_s}{6} L_b-
\frac{4n_F}{3} L_f
+\frac{53}{6} -\frac{N_s}{3}+\frac{4n_F}{3}
+\frac{3}{2}h^2-3t^2\biggr)\biggr]\, , \nonumber\\\\
g_{3,\mu=0}^2 & = & g^2(\mu)T\biggl[1+\frac{g^2}{16\pi^2}\biggl(
\frac{44-N_s}{6} L_b-
\frac{4n_F}{3} L_f+\frac{2}{3}\biggr)\biggr] \, ,
\end{eqnarray}
where $h\equiv m_H/m_W$ and $t\equiv m_t/m_W$, with $m_H$, $m_W$ and $m_t$ denoting the physics Higgs, $W$-boson and top quark masses, respectively.

The scalar mass parameter in the presence of finite chemical potential is given by
\begin{eqnarray}
m_3^2(\bar{\mu}) = m_{3,\mu=0}^2(\bar{\mu}) + \Delta m_3^2 
\end{eqnarray}
where the correction due to finite chemical potential is given by
\begin{eqnarray}
\Delta m_3^2 & = & \frac{g_Y^2(\bar{\mu})}{12\pi^2}\mu_B^2 - \frac{3g_Y^2}{16\pi^2}\left(\nu^2 - \frac{\lambda T^2}{2} - \frac{3g^2T^2}{16} 
- \frac{g_Y^2T^2}{4}\right){\cal A}\left(\frac{\mu_B}{3}\right)
+ \frac{g_Y^2\mu_B^2}{64\pi^4}\left[\frac{3}{4}g^2L_b(\bar{\mu}) -g_Y^2 \right.\nonumber\\
&& \left. \times \left(L_f(\bar{\mu})-{\cal A}\left(\frac{\mu_B}{3}\right)\right)\right]  -\Bigg[\left(9g_Y^4+\frac{9}{2}g_Y^2g^2+16g_Y^2g_s^2\right){\cal A}\left(\frac{\mu_B}{3}\right)
-\left(9g_Y^4+\frac{9}{4}g_Y^2g^2 \right.\nonumber\\
&&\left. -18\lambda g_Y^2
-16g_Y^2g_s^2-\frac{27}{4}g^4\right)16{\cal B}\left(\frac{\mu_B}{3}\right)\Bigg]\frac{T^2}{128\pi^2}  +\left(9g_Y^4+\frac{9}{4}g_Y^2g^2-18\lambda g_Y^2-16g_Y^2g_s^2 \right. \nonumber \\
& & \left.-\frac{27}{4}g^4\right)\frac{i\mu_B T}{48\pi^3}
\ln\left(\frac{\Gamma\left(\frac{1}{2}-\frac{i\mu_B}{6\pi T}\right)}{\Gamma\left(\frac{1}{2}+\frac{i\mu_B}{6\pi T}\right)}\right) +\Bigg[9g_Y^2g^2 - 6\left(3g_Y^4-8g_Y^2g_s^2\right)L_b(\bar{\mu})+9g_Y^4L_f(\bar{\mu})\nonumber \\
& &
+\left(\frac{9}{2}g_Y^2g^2 + 16g_Y^2g_s^2\right)\left(4\ln\,2 - 1\right)  +\left(9g_Y^4+\frac{9}{4}g_Y^2g^2-18\lambda g_Y^2-16g_Y^2g_s^2-\frac{27}{4}g^4\right)4\gamma_E \nonumber \\
& &- \left(9g_Y^4+\frac{9}{2}g_Y^2g^2+16g_Y^2g_s^2\right){\cal A}\left(\frac{\mu_B}{3}\right)\Bigg]
\frac{\mu_B^2}{1152\pi^4}  -\frac{3}{4}g^4\sum_{i=1}^{n_f}\left[\frac{T^2}{8\pi^2}{\cal B}\left(\mu_{L_i}\right)+\frac{i\mu_{L_i} T}{16\pi^3} \right. \nonumber \\
& & \left. \times
\ln\left(\frac{\Gamma\left(\frac{1}{2}-\frac{i\mu_{L_i}}{2\pi T}\right)}{\Gamma\left(\frac{1}{2}+\frac{i\mu_{L_i}}{2\pi T}\right)}\right)
+\frac{\mu_{L_i}^2}{32\pi^4}\gamma_E\right]\, .
\end{eqnarray}
The function ${\cal B}(\mu)$ is defined as
\begin{eqnarray}
{\cal B}(\mu) & \equiv & \zeta'\left(-1,\frac{1}{2}+\frac{i\mu}{2\pi T}\right) + \zeta'\left(-1,\frac{1}{2}-\frac{i\mu}{2\pi T}\right)  - 2\zeta'\left(-1,\frac{1}{2}\right)
\end{eqnarray}
with $\zeta(z,q)$ being the generalised zeta function and the functions $L_b(\bar{\mu})$ and $L_f(\bar{\mu})$ are given by
\begin{eqnarray}
L_b(\bar{\mu}) & = & \ln\frac{\bar{\mu}^2}{T^2} - 2\ln 4\pi + 2\gamma_E  \\
L_f(\bar{\mu}) & = & \ln\frac{\bar{\mu}^2}{T^2} - 2\ln \pi + 2\gamma_E.
\end{eqnarray}
with $\bar{\mu}$ denoting the renormalisation scale in the $\bar{\text{MS}}$ scheme. The scalar mass parameter for zero chemical potential is given by
\begin{eqnarray}
m_{3,\mu=0}^2(\bar{\mu}) & = & -\tilde{\nu}^2
+T\biggl(\frac{1}{2}\lambda_3+\frac{3}{16}g_3^2+\frac{1}{16}g_3'^2+
\frac{1}{4}\tilde{g}_Y^2\biggr)
+
\frac{T^2}{16\pi^2} \biggl[
g^4\biggl(\frac{137}{96}+\frac{3n_F}{2} \ln{2}+\frac{n_F}{12}\biggr)+
\frac{3}{4}\lambda g^2 \biggr]
\nonumber \\
& &+ 
\frac{1}{16\pi^2}
\biggl(\frac{39}{16}g_3^4+12h_3g_3^2-6h_3^2+
9\lambda_3g_3^2-12\lambda_3^2\biggr)
\biggl(\ln\frac{3 T}{\mu}+c\biggr)\, ,
\label{m32}
\end{eqnarray}
with
\begin{eqnarray}
\tilde{\nu}^2 & = & \nu^2(\bar{\mu})\biggl\{1- \frac{3}{4}\frac{g^2}{16\pi^2}
\biggl[\biggl(
h^2-3\biggr)L_b+2 t^2 L_f\biggr]\biggr\} \, , \\
\tilde{g}_Y^2 & = & T g_Y^2(\bar{\mu})\biggl\{1-\frac{3}{8}
\frac{g^2}{16\pi^2}\biggl[\biggl(6 t^2-
6-\frac{64}{3}s^2\biggr)L_f\nonumber \\
& + & 2+28 \ln{2}-12h^2\ln 2+
8 t^2\ln 2-\frac{64}{9}s^2(4\ln 2-3)\biggr]\biggr\} \, .
\end{eqnarray}
Here $s$ is defined as $s\equiv g_s/g$. At the two-loop level (ignoring contributions
of the order, $g^{\prime 3}\sim g^{9/2}$) the coefficient of the new term arising due to finite potential is given by
\begin{eqnarray}
\kappa_1 & = & -\frac{i\pi}{3}g'T^{5/2}\Bigg[\left(1 - \frac{9g^2}{64\pi^2}\right)\sum_{i=1}^{n_f} \frac{\mu_{L_i}}{\pi T}\left(1+\left(\frac{\mu_{L_i}}{\pi T}\right)^2\right) \nonumber \\
& & -\left(1-\frac{5g_Y^2}{32\pi^2}-\frac{9g^2}{64\pi^2}-\frac{g_s^2}{2\pi^2}\right)\frac{\mu_B}{\pi T}\left(1+\frac{1}{9}\left(\frac{\mu_B}{\pi T}\right)^2\right)\Bigg] \, .  
\end{eqnarray}
We also take into account the running of all the coefficients by using one-loop 
\section{Entropy dilution}
\label{sec:Entropy dilution}
As we have discussed above, if we want the sphaleron freeze-in mechanism to produce the correct baryon asymmetry observed today, then to be consistent with the constraints from BBN the large [$\mathcal{O}(1)$] primordial lepton asymmetry should be diluted somewhere between the electroweak symmetry breaking and before the onset of neutrino oscillations (at around $T\sim 10$ MeV). It is important to stress that the entropy dilution is not necessary for a first-order QCD transition as discussed in the following section. Therefore, the entropy dilution can in principle occur after the electroweak symmetry breaking once the baryon asymmetry is generated and before the QCD phase transition, leading to the correct baryon asymmetry. 
Such an injection of entropy in the thermal bath can occur via a late-time decaying state.  The key feature for the late-time decaying state is that it should redshift like matter such that it can dominate the Universe's energy budget following radiation domination providing late-time matter domination~\cite{Evans:2019jcs}.
Furthermore, such a state must decay mostly into the SM states in a thermal bath providing an era of late-time reheating. The decay rate of the late-time decaying state is governed by the mass of such state $\Gamma_m$, and the co-moving energy stored in such a state, $\Phi_m \equiv \rho_{m} a^3$, where $a$ is the scale factor. 

The energy of the universe can be separated into two contributions that are constant during purely adiabatic expansion; the contribution $\Phi_m \equiv \rho_{m} a^3$ from the late-time decaying state, and the contribution from radiation, $\Phi_R\equiv \rho_R a^4$. The Hubble expansion rate is therefore
 \begin{equation}
H = \frac{\dot a}{a} = \frac{1}{M_{\rm pl}} \sqrt{\frac{8 \pi}{3} \left( \frac{\Phi_R}{a^4} +\frac{\Phi_{m}}{a^3} \right)},
\label{eq:Hub}
 \end{equation}
where $M_{\rm pl}$ is the Planck mass. 

\subsection{Matter-radiation equality} Given how these two different densities scale, at some time $t=t_{\rm eq}$, the matter and radiation energy densities would become equal. We will denote this by  $\rho_{{m},\rm{eq}}=\rho_{R,\rm{eq}}$, with $\rho_{{ m},\rm{eq}}\equiv \rho_{m}(t_{\rm eq})$, $\rho_{R,\rm{eq}}\equiv \rho_{R}(t_{\rm eq})$. Note that we can always define the scale factor at $t_{\rm eq}$ to be $a_{\rm eq}\equiv 1$, such that $\Phi_{m,\rm{eq}} = \Phi_{R,\rm{eq}}$. The temperature $T_{\rm eq}$ of the SM plasma at the time of matter-radiation equality, $t_{\rm eq}$, when $\rho_{{m},\rm{eq}}=\rho_{R,\rm{eq}}$, is given by
 \begin{equation}
 T_{\rm eq} = \left( \frac{30 \, \Phi_m}{ \pi^2 g_{*}(T_{\rm eq})} \right)^{\frac 14}  \approx 1.32 \left( \frac{\Phi_m}{g_{*}(T_{\rm eq})}\right)^{\frac 14},
 \label{eq:Teq}
 \end{equation}
 where $g_{*}(T)$ is the effective number of relativistic degrees of freedom (d.o.f.)\ in the SM at temperature $T$.
 
 \subsection{Adiabatic matter domination} Following the start of domination of matter, there is potentially an adiabatic matter domination phase before the late-time decaying state decays in the bulk (leading to non-adiabatic matter domination at $t_{\rm NA} = \Gamma_m^{-1}$). During this phase $\Phi_m\simeq \Phi_{m,\rm{eq}}=$const, which leads to
  \begin{equation}
 \Phi_R(a) = \Phi_{m,\rm{eq}} + \frac{2}{5}  \sqrt{\frac{3}{8\pi}}  M_{pl} \Gamma_m \sqrt{\Phi_{m,\rm{eq}}}  \left( a^{\frac 52} -1 \right).
 \label{eq:Ra}
 \end{equation}

\subsection{Non-adiabatic matter domination}
Once the second term in \eqref{eq:Ra} is comparable in size to $\Phi_{m,\rm{eq}}$, the evolution enters a non-adiabatic phase. We equate the two terms to define this transition temperature, $T_{\rm NA}$. For $a_{\rm NA}\gg a_{\rm eq}=1$, this is given by
  \begin{equation}
T_{\rm NA} = T_{\rm eq} \left( \frac{g_*(T_{\rm eq})}{g_*(T_{\rm NA})}\right)^{\frac 13} \left(\frac{2}{5}  \sqrt{\frac{3}{8\pi}} \frac{M_{pl} \Gamma_m}{\sqrt{\Phi_{m,i}}} \right)^{\frac25} 
\approx 0.59 \left( \frac{g_*^{\frac1{4}}(T_{\rm eq})}{g_*(T_{\rm NA})} \right)^{\frac13}  \left( M_{pl} \Gamma_m \Phi_{m,i}^{\frac 18}\right)^{\frac25},
\label{eq:TNA}
 \end{equation}
where we used that up to this point the evolution is adiabatic, and therefore $g_*(T_{\rm eq}) T_{\rm eq}^3 a_{\rm eq}^3 =g_*(T_{\rm NA})T_{\rm NA}^3 a_{\rm NA}^3$. 

\subsection{Reheating}
Once the late-time decaying state decays away in bulk at temperature $T_d$, the Hubble expansion takes over at around $\Gamma_m \sim H$, and adiabatic expansion resumes.  This transition happens at the decay temperature of the late-time decaying state $T_d$ and defines the reheat temperature of the universe \cite{Kofman:1997yn}, 
\begin{equation}
T_d\equiv T_{\rm RH} = \left( \frac{90}{8\pi^3 g_*(T_{\rm RH})}\right)^{\frac14} \sqrt{\Gamma_m M_{pl}} \approx 0.78  g_*^{-\frac14}(T_{\rm RH}) \sqrt{\Gamma_m M_{pl}} .
\label{eq:TRH}
\end{equation}

\subsection{Entropy dilution of asymmetry}
In the event of a late-time decay and reheating as described above, any frozen-out lepton or baryon asymmetry decoupled from the SM plasma (such that there is no longer an efficient energy transfer between the SM and the decoupled asymmetry, even though a chemical equilibrium is still maintained) undergoes a dilution. This is because the late-time decaying field decays dominantly into the thermal bath SM particles (still in equilibrium), and then the entropy dumped into the SM plasma by the late-time decaying state will lead to an entropy dilution of the decoupled species. We stress that in the definition of the lepton asymmetry abundance $Y_{L}\equiv n_L-n_{\bar{L}}/s$, $s$ corresponds to the entropy of the thermal bath (This is not to be confused with the separately conserved entropy of the decoupled lepton asymmetry abundance $s_L$, which scales with the decoupled temperature (due to a non-efficient energy transfer between the SM and the decoupled asymmetry) of the frozen out abundance.). Therefore, if the thermal bath undergoes reheating due to a late-time decay of heavy states then the thermal bath temperature and consequently also the thermal bath entropy will change leading to a dilution of any frozen-out asymmetry.   

In our case such a dilution should occur at quite late times after the electroweak phase transition and before the onset of neutrino oscillations (at around $T\sim 10$ MeV).  

We can define the entropy dilution factor $\Delta$ as
\begin{equation}
{Y_L}\equiv \frac{1}{\Delta} Y_L^{\slashed{\Delta}_s}.
\end{equation}
Here the left-hand side represents the lepton asymmetry abundance in the presence of the entropy production due to the decay of a late-time decaying state. The abundance of the
lepton asymmetry on the right-hand side represents the case without the entropy production. Now, there are two possible cases of interest depending on when the lepton asymmetry freeze-out temperature ($T_{\rm fo}$) : (a) $T_{\rm fo}>T_{\rm eq}>T_{\rm RH}$ and (b) $T_{\rm eq} > T_{\rm fo}>T_{\rm RH}$; for which we provide approximate solutions below.

In case (a), assuming that the lepton asymmetry is already decoupled from the SM plasma when matter domination begins at $T_{\rm eq}$, the entropy dilution after the reheating of the SM plasma is given by
\begin{equation}
\Delta= \left(Y_L \right)^{-1}  Y_L^{\slashed{\Delta}_s}\simeq \left(\frac{s}{\rho}\right)_{\rm RH} \left(\frac{\rho}{n_{\Delta L}}\right)_{eq} \left(\frac{n_{\Delta L}}{s}\right)_{\rm eq}\simeq  T_{\rm eq}/T_{\rm RH}
\end{equation}
where we use $Y_L^{\slashed{\Delta}_s}\simeq (Y_L)_{\rm eq}$ assume that most of the late-time decaying state decays around $T_{\rm RH}$. We note that the subscript to the parentheses denotes the relevant time where the quantity in the parentheses is to be evaluated. 

In case (b) the lepton asymmetry decouples from the thermal equilibrium
when the late-time decaying state is dominating the energy budget of the Universe. In this case, we need to know the dependence of the lepton asymmetry density on the Hubble parameter
at the decoupling of the lepton asymmetry. Since the decoupling takes place when the rates of all the lepton number violating (LNV) processes become comparable to the expansion rate
\begin{equation}
n_{\Delta L,fo} \;\sim\;\frac{H_{fo}}{\langle \sigma v \rangle}_{\rm LNV}.
\end{equation}
The entropy dilution in this scenario can be estimated as
\begin{equation}
\Delta= \left(Y_L \right)^{-1}  Y_L^{\slashed{\Delta}_s} \simeq  \left(\frac{s}{\rho}\right)_{\rm RH} \left(\frac{\rho}{n_{\Delta L}}\right)_{\rm fo} \left(\frac{n_{\Delta L}}{s}\right)_{{\rm fo}}^{\slashed{\Delta}_s}
	     \simeq \frac{T_{\rm fo}^{\slashed{\Delta}_s}}{ T_{\rm RH}} \left(\frac{H_{\rm fo}}{H_{\rm fo}^{\slashed{\Delta}_s}}\right),
\label{delta}	     
\end{equation}
where we have used $(\rho)_{\rm fo}= H_{\rm fo}^2 M_{pl}^2$, $(n_{\Delta L})_{\slashed{\Delta}_s} \sim H_{\rm fo}^{\slashed{\Delta}_s}/\langle \sigma v \rangle_{\rm LNV}$, $(s)_{\slashed{\Delta}_s}\sim(H_{\rm fo}^{\slashed{\Delta}_s})^2 M_{pl}^2/T_{\rm fo}$.

In the case when $T_{\rm fo} >T_{\rm NA}$ ($T_{\rm NA}$ corresponds to the temperature when the new radiation from the decaying late time state starts to dominate over the preexisting radiation component of the Universe) we have 
\begin{equation}
\Delta \sim  \frac{ T_{\rm NA}^{5/2} T_{\rm fo}^{3/2}}{T_{\rm RH}^3 T_{\rm fo}^{\slashed{\Delta}_s}}\sim  \frac{ T_{\rm eq}^{1/2} T_{\rm fo}^{3/2}}{T_{\rm RH} T_{\rm fo}^{\slashed{\Delta}_s}},
\end{equation}
where we have used $H_{\rm fo}\propto T_{\rm NA}^{5/2} T_{\rm RH}^{-2} T_{\rm fo}^{3/2}$, with  $T_{\rm eq} \sim T_{\rm NA}^5/T_{\rm RH}^4$ and $H_{\rm fo}^{\slashed{\Delta}_s} \propto (T_{\rm fo}^{\slashed{\Delta}_s})^2$.

On the other hand, for the case $T_{\rm fo} <T_{\rm NA}$ we have 
\begin{equation}
\Delta \sim \frac{T_{\rm fo}^4}{T_{\rm RH}^3 T_{\rm fo}^{\slashed{\Delta}_s}}.
\end{equation}

\section{Note on equilibrium}
\label{app:Note on equilibrium}

QCD transitions are calculated assuming one has an infinite amount of time, to make the problem tractable. This means that the system always has enough to equilibrate and the nucleation temperature is the same as the critical temperature - that is the highest temperature at which it is energetically favorable to change the ground state of the system. In the case of a scalar field theory, the potential is double-welled at the critical temperature and if the system was given enough time to equilibrate, the transition would happen at the critical temperature. Note that the value of the potential in each minimum is the value of the pressure in each phase. So the pressure difference vanishes at the critical temperature. In practice, the system cools until the nucleation rate grows large enough compared to the expansion rate of the Universe such that there is at least one critical-size bubble per Hubble volume. 
In the case of QCD, we have no scalar potential. However, the critical temperature is still defined as the moment when the pressure difference vanishes. If the pressure difference vanishes, there is no change in the yield because there is no vacuum energy converting to a dump of particles in the plasma. So if it is a good approximation that the phase transition occurs near the critical temperature, it is a good approximation that we are always in thermal equilibrium as the change in the yield will be small. So it is worth discerning how much supercooling there is likely to be.
There is no equivalent for the bounce action in QCD, but we can rely on classical nucleation theory for insight. Classical nucleation theory gives identical results to the full scalar theory in the thin wall approximation, which is valid at or near the critical temperature. Consider the energy of a bubble 
\begin{equation}
    E = - \frac{4 \pi }{3} R^3 \Delta p + 4 \pi R^2 \sigma
\end{equation}
where $\Delta p$ is the pressure difference and $\sigma $ is the surface tension between the phases. Such a bubble has an extrema at $R= 2\sigma  / \Delta p$. Below this critical radius, the bubble collapses under the surface tension, above it the pressure wins and the bubble expands. Since it is exponentially costly to nucleate a bubble, the phase transition is dominated by bubbles infinitesimally larger than the critical radius. The nucleation rate in classical nucleation theory is set by
\begin{eqnarray}
    \Gamma \sim e^{-E_C/T} \sim e^{-\frac{16 \pi \sigma ^3}{3 \delta p ^2T}} \ .
\end{eqnarray}
The pressure difference grows as we supercool below the critical temperature. How much we supercool will depend on the surface tension as this will determine when we nucleate. For the sake of illustration, suppose the pressure difference scales as
\begin{equation}
    \Delta p \sim \delta g_{\ast}\Delta T^4 
\end{equation}
where $\Delta T$ is $T_C-T_*$ aka the amount of temperature we have supercooled below the critical temperature and $g_\ast \sim 40$ is the change in the relativistic d.o.f due to the phase transition. Next suppose the surface tension scales as,
\begin{equation}
    \sigma \sim \epsilon T_C^3 \ .
\end{equation}
For our lowest value of the critical temperature, we consider that $\sigma \sim 5 {\rm fm}^{-2} \ {\rm MeV}$ implying $\epsilon \sim 10^{-3}$\cite{Gao:2016hks}. The nucleation temperature is when $E_C/T\sim 140$ which gives $\Delta T/T\sim 3\%$ supercooling. This is enough to give a boost to our estimate in the gravitational waves, but this is unlikely to make a large change in the yield. 

Let us conclude by addressing a potential point of confusion. Electroweak baryogenesis requires a departure from equilibrium which is usually taken as being equivalent to demanding a first order electroweak phase transition. So how can we say that we have a limited departure from equilibrium if we have a first order QCD transition? In electroweak baryogenesis, we set up a Boltzmann equation for the baryon number which require the following hierarchy of scales
\begin{equation}
    \tau _{\rm sph, \ broken}<<\tau _{\rm PT} << (\tau _{sph, symm}, \tau _{CPV})
\end{equation}
If we did not have the first hierarchy, the sphalerons inside the bubble would wash out the baryon asymmetry. So baryogenesis always requires some minimal amount of supercooling and therefore some amount of departure from equilibrium.

\section{Matching Standard Model parameters to observables}\label{app:matching}
We use SARAH~\cite{Staub:2008uz} to calculate the evolution of running couplings matched at the Z pole. Note that our conventions differ from SARAH's in that the Higgs quartic is rescaled by a factor of 2 and the U(1) hypercharge gauge coupling constant is rescaled by a factor of $\sqrt{5/3}$. Our effective potential and the RGEs we use are only functions of the most numerically important couplings ${\lambda, \mu ^2 , g_i, y_t}$ whose values at the Z-pole we give in Tab.~\ref{tab:match}.

A recent work \cite{Huang:2020hdv} gave multiloop matching at the Z pole for the couplings we are interested in and we follow their analysis here. Special attention is given to the top mass due to the slow convergence of perturbation theory - both the Yukawa coupling and the strong coupling are O(1). The relationship between the top mass and the top Yukawa considering only QCD corrections is known to four loops \cite{Marquard:2016dcn}
\begin{equation}
    M_t = \frac{y_t(m_t) v}{\sqrt{2}} \left(1+ 0.4244 \alpha _s +0.8345 \alpha _s^2 +2.37 \alpha _s^3 +(8.615 \pm 0.017) \alpha _s^4 \right) \ .
\end{equation}
\begin{table}[t]
    \centering
    \begin{tabular}{c|c}
        Parameter & Value \\
\hline         $\lambda $ & 0.13947 \cite{Croon:2020cgk} \\
$\mu ^2$ (GeV) & -8434 \cite{Croon:2020cgk} \\
    $g_1$ & 0.357254 \cite{Huang:2020hdv} \\
    $g_2$ & 0.651 \cite{Huang:2020hdv} \\
        $g_3$ & 1.2104 \cite{Huang:2020hdv} \\
$y_t$ & 0.95367 \cite{Huang:2020hdv}
    \end{tabular}
    \caption{Central value of parameters with references referring to the methods used to derive the values in question. }
    \label{tab:match}
\end{table}
Here the mass on the left-hand side, $M_t$, is the pole mass for the top. Throughout we will capitalize pole masses.
However, this is insufficient to match as the 1 loop non-QCD corrections become numerically important for the large values of the scale we consider in this work. The full self-energy of the top at one loop can be written as \cite{Croon:2020cgk}
\begin{equation}
    \frac{1}{2}y_t^2v^2 = M_t^2\left( 1+ 2 {\rm Re} \left(  \Sigma _v (M_t^2) + \Sigma _s (M_t^2) \right) \right) 
\end{equation}
where
\begin{align}
\Sigma_v(M^2_t) + \Sigma_s(M^2_t) &=
    \frac{3}{16} \frac{g^2_2}{(4\pi)^2} \bigg( -2 - 4 \frac{1}{h^2} - 2 h^2 - \frac{256}{9}s^2 + 2 t^2 + 16 \frac{t^4}{h^2}
    \nonumber \\ &
    - \frac{2}{27} \Big( 39 - \frac{64}{z^2} + 25 z^2 + 18\frac{z^4-1}{h^2} \Big)
     \nonumber \\ &
    + \Big(4 h^2 - \frac{8}{3}t^2 + \frac{4}{3} t^2 \frac{2t^2 + h^2}{t^2-h^2} \Big) \ln(h) - \frac{8}{9} \Big( -9 \frac{z^4}{h^2} + 4 \frac{(4-5z^2+z^4)}{t^2-z^2} \Big) \ln(z)
      \nonumber \\ &
    + \Big( \frac{128}{3}s^2 -32 \frac{t^4}{h^2} - \frac{4}{3} \frac{t^2(2t^2 + h^2)}{t^2-h^2} - \frac{32}{9} \frac{(z^2-1)(t^2-4) }{z^2-t^2} \Big) \ln(t)
     \nonumber \\ &
    + \frac{2}{3}\Big(4t^2 -h^2  \Big) F(M_t,M_t,M_h)
    + \frac{2}{3}\frac{(t^2+2)(t^2-1)}{t^2} F(M_t,M_W,0)
      \nonumber \\ &
    - \frac{2}{27}\Big( \frac{64 - 80z^2 + 7z^4}{z^2}
     + \frac{32-40z^2 + 17z^4}{t^2} \Big) F(M_t,M_t,M_Z)
      \nonumber \\ &
    + \bigg[ 2\Big(-6 \frac{1}{h^2} - h^2 -\frac{32}{3}s^2 + t^2 + 8 \frac{t^4}{h^2} \Big)
      \nonumber \\ &
    - \frac{4}{9}\frac{(z^2-1)(9+4h^2+9z^2)}{h^2} \bigg] \ln\Big( \frac{Q^2}{M^2_W} \Big) \bigg)
    \nonumber \\ &
    - 64 c_6 \frac{M^2_W}{g^4_0} \bigg[1 - \ln\Big( \frac{Q^2}{M^2_h} \Big) \bigg] \bigg)
    \;,\\[3mm]
\frac{\delta g^{2}}{g^2_0} &=
    \frac{1}{(4\pi)^2} g^2_0  \bigg(
    - \frac{257}{72}
    - \frac{1}{24}h^2
    + \frac{20}{9} N_f
    + \frac{1}{4}t^2 
    - 2\ln(t)
   \nonumber \\ &
    + \frac{1}{12}(12 - 4 h^2 + h^4) F(M_W,M_h,M_W)
    - \frac{(t^2+2)(t^2-1)}{2} F(M_W,M_t,0)
   \nonumber \\ &
    - \frac{33}{4}F(M_W,M_W,M_W)
    + \Big(\frac{4}{3} N_f -\frac{43}{6} \Big)\ln\Big( \frac{Q^2}{M^2_W} \Big)
    \bigg)\;
\end{align}
where $\{ h,t,z\} =\{ M_h/M_W, M_t/M_W, M_Z/M_W \}$, $Q$ is the $\overline{MS}$ RG scale, and the loop function, $F(k,m_1,m_2)$, is given in Ref.~\cite{Kajantie:1995dw}.  An accurate calculation requires we can extract the relationship between the observed top mass and the Yukawa including at least 1 loop correction for everything but QCD to which we go to four loops (as shown in ref \cite{Huang:2020hdv}). For the top mass, we take a central value of the top quark to be $172.4 \pm 0.7$ GeV.

For the gauge couplings, we use the parameters derived in Ref. \cite{Huang:2020hdv} whose 2 loop matching derives from refs. \cite{Degrassi:2014sxa,Martin:2015rea}.

For the Higgs sector parameters, we follow Ref.~\cite{Croon:2020cgk}
\begin{eqnarray}
\mu _h ^2 &=& - \frac{1}{2} M_h ^2 \left( 1+ \frac{{\rm Re}\Pi _h (M_h ^2)}{M_h ^2}\right) \\
\lambda &=& \frac{1}{8} g_2^2 \frac{M_h^2}{M_W^2} \left(1 + \frac{{\rm Re} \Pi _h (M_h^2)}{M_h^2} - \frac{{\rm Re} \Pi _W (M_W^2)}{M_W^2} \right)
\end{eqnarray}
where the self-energies are given by
\begin{align}
\Pi_h(M^2_h) &=
    \frac{3}{8} \frac{g^2_0 M^2_h}{(4\pi)^2} \bigg(
    - \frac{4}{3}
    - 8 \frac{1}{h^2}
    - 2h^2
    + 16 \frac{t^4}{h^2}
    - \frac{2}{3} z^2
    - 4\frac{z^4}{h^2}
      \nonumber \\ &
    + 3h^2 F(M_h,M_h,M_h)
    + 4t^2\Big( 1 - 4 \frac{t^2}{h^2} \Big) F(M_h,M_t,M_t)
      \nonumber \\ &
    + \frac{2}{3}\frac{h^4 - 4h^2 + 12}{h^2} F(M_h,M_W,M_W)
      \nonumber \\ &
    + \Big( \frac{1}{3} \frac{1}{h^2} - \frac{4}{3}z^2 + 4\frac{z^4}{h^2} \Big) F(M_h,M_Z,M_Z)
      \nonumber \\ &
    - 2h^2 \ln(h)
    - 8t^2 \ln(t)
    + \Big( -\frac{2}{3}h^2 + 4 z^2 \Big) \ln(z)
      \nonumber \\ &
    + \Big( -4 +2h^2 + 4 t^2 - 2 z^2 \Big) \ln\Big( \frac{Q^2}{M^2_W} \Big)
      \nonumber \\ &
    + 64 c_6 \frac{M^2_W}{g^4_0 h^4} \bigg[ -2 + 12 t^4 - z^4 +  3 h^4 F(M_h,M_h,M_h)
    - 6 h^4 \ln(h)
      \nonumber \\ &
    - 24 t^4 \ln(t)
    + 6z^4 \ln(z)
    + \Big( -2 + h^4 + 4 t^4 - z^4 \Big) \ln\Big( \frac{Q^2}{M^2_W} \Big) \bigg]
      \nonumber \\ &
    + 3072 c_6^{2} \frac{M^4_W}{g^4_0 h^2} \bigg[ -1 + F(M_h,M_h,M_h) \bigg]\bigg)
    \;,
\end{align}
\begin{align}
\Pi_W(M^2_W) &=
    \frac{3}{8} \frac{g^2_0 M^2_W}{(4\pi)^2} \bigg(
    - \frac{212}{9}
    - \frac{8}{3}\frac{1}{h^2}
    - \frac{22}{9} h^2
    + \frac{4}{27} (40 N_f - 17)
    - \frac{4}{3}t^2
    + 16 \frac{t^4}{h^2}
    + \frac{14}{9}z^2
    - \frac{4}{3} \frac{z^4}{h^2}
      \nonumber \\ &
    + \frac{4h^2 (h^2-2)}{h^2-1} \ln(h)
    - 8\Big( \frac{2}{3} - t^2 + 4\frac{t^4}{h^2} \Big) \ln(t)
    + 4\Big(2 \frac{z^4}{h^2} - \frac{z^4 - 4 z^2 - 8}{z^2-1} \Big) \ln(z)
     \nonumber \\ &
    + \frac{2}{9}\Big( 12 - 4 h^2 + h^4 \Big) F(M_W,M_h,M_W)
    - \frac{4}{3}(t^2+2)(t^2-1) F(M_W,M_t,0)
      \nonumber \\ &
    - \frac{32}{3}\frac{z^2-1}{z^2} F(M_W,M_W,0)
    + \frac{2}{9}\frac{(z^4 + 20 z^2 + 12)(z^2-4)}{z^2} F(M_W,M_W,M_Z)
      \nonumber \\ &
    + 2\bigg[-1 + \frac{2}{h^2} + \Big( -\frac{59}{9} - 6 \frac{1}{h^2} - h^2 + \frac{16}{9} N_f - 2 t^2 + 8 \frac{t^4}{h^2} \Big) + z^2 - 2\frac{z^4}{h^2} \bigg] \ln\Big( \frac{Q^2}{M^2_W} \Big)
    \nonumber \\ &
    - \frac{8}{9}\pi i (4 N_f - 3)
    - 64 c_6 \frac{M^2_W}{g^4_0} \bigg[1 - \ln\Big( \frac{Q^2}{M^2_h} \Big) \bigg]  \bigg)
    \;, 
\end{align}

\bibliographystyle{utphys}
\bibliography{mm}
\end{document}